\documentclass[aj,twocolumn]{aastex62}

\usepackage{placeins}
\usepackage{comment}

\newcommand{\thisstar}{HIP 94235}
\newcommand{\thisstarb}{HIP 94235 b}

\newcommand{\kms}{\ensuremath{\rm km\,s^{-1}}}
\newcommand{\ms}{\ensuremath{\rm m\,s^{-1}}}

\newcommand{\rstar}{\ensuremath{R_\star}}

\newcommand{\mearth}{\ensuremath{M_\earth}}

\newcommand{\rpl}{\ensuremath{R_{p}}}
\newcommand{\mpl}{\ensuremath{M_{p}}}

\submitjournal{AJ}

\shorttitle{\thisstarb{} - A Mini-Neptune in AB Dor MG}
\shortauthors{Zhou, Wirth, Huang et al.}

\begin{document}

\title{A Mini-Neptune from \emph{TESS} and \emph{CHEOPS} Around the 120 Myr Old AB Dor member \thisstar{}}

\correspondingauthor{George Zhou}
\email{george.zhou@usq.edu.au}

\author[0000-0002-4891-3517]{George Zhou} 
\affil{University of Southern Queensland, Centre for Astrophysics, West Street, Toowoomba, QLD 4350 Australia}

\author[0000-0003-1656-011X]{Christopher P. Wirth}
\affiliation{Harvard University, Cambridge, MA 02138, USA.}
\affiliation{Center for Astrophysics \textbar{} Harvard \& Smithsonian, 60 Garden St., Cambridge, MA 02138, USA.}

\author[0000-0003-0918-7484]{Chelsea X. Huang}  
\affil{University of Southern Queensland, Centre for Astrophysics, West Street, Toowoomba, QLD 4350 Australia}

\author[0000-0002-8400-1646]{Alexander Venner} 
\affiliation{Aberdeen, UK}

\author{Kyle Franson} 
\affil{Department of Astronomy, The University of Texas at Austin, TX 78712, USA}

\author[0000-0002-8964-8377]{Samuel N.~Quinn} 
\affiliation{Center for Astrophysics \textbar{} Harvard \& Smithsonian, 60 Garden St., Cambridge, MA 02138, USA.}

\author[0000-0002-0514-5538]{L. G. Bouma} 
\altaffiliation{51 Pegasi b Fellow}
\affiliation{Cahill Center for Astrophysics, California Institute of Technology, Pasadena, CA 91125, USA}

\author[0000-0001-9811-568X]{Adam L. Kraus}  
\affil{Department of Astronomy, The University of Texas at Austin, Austin, TX 78712, USA}

\author[0000-0003-3654-1602]{Andrew W.~Mann} 
\affiliation{Department of Physics and Astronomy, The University of North Carolina at Chapel Hill, Chapel Hill, NC 27599, USA}

\author[0000-0003-4150-841X]{Elisabeth.R.~Newton} 
\affiliation{Department of Physics and Astronomy, Dartmouth College, Hanover, NH 03755, USA}

\author{Diana Dragomir} 
\affiliation{Department of Physics and Astronomy, University of New Mexico, 210 Yale Blvd NE, Albuquerque, NM 87106, USA}

\author[0000-0002-8091-7526]{Alexis Heitzmann} 
\affil{University of Southern Queensland, Centre for Astrophysics, West Street, Toowoomba, QLD 4350 Australia}

\author[0000-0001-6508-5736]{Nataliea Lowson}
\affil{University of Southern Queensland, Centre for Astrophysics, West Street, Toowoomba, QLD 4350 Australia}

\author{Stephanie T. Douglas}  
\affiliation{Department of Physics, Lafayette College, 730 High St., Easton, PA 18042, USA}

\author[0000-0002-1357-9774]{Matthew Battley} 
\affiliation{Department of Physics, University of Warwick, Gibbet Hill Road, Coventry CV4 7AL, UK}
\affiliation{Centre for Exoplanets and Habitability, University of Warwick, Gibbet Hill Road, Coventry CV4 7AL, UK}

\author[0000-0003-2851-3070]{Edward Gillen} 
\altaffiliation{Winton Fellow}
\affiliation{Astronomy Unit, Queen Mary University of London, Mile End Road, London E1 4NS, UK}
\affiliation{Astrophysics Group, Cavendish Laboratory, J.J. Thomson Avenue, Cambridge CB3 0HE, UK}


\author{Amaury Triaud} 
\affiliation{School of Physics \& Astronomy, University of Birmingham, Edgbaston, Birmingham, B15 2TT, UK}


\author[0000-0001-9911-7388]{David W.~Latham}  
\affiliation{Center for Astrophysics \textbar{} Harvard \& Smithsonian, 60 Garden St., Cambridge, MA 02138, USA.}

\author[0000-0002-2532-2853]{Steve~B.~Howell}  
\affiliation{NASA Ames Research Center, Moffett Field, CA 94035, USA}

\author[0000-0001-8732-6166]{J. D. Hartman} 
\affiliation{Department of Astrophysical Sciences, Princeton University, 4 Ivy Lane, Princeton, NJ 08540, USA}

\author[0000-0003-2053-0749]{Benjamin M.~Tofflemire}  
\affil{Department of Astronomy, The University of Texas at Austin, Austin, TX 78712, USA}


\author[0000-0001-9957-9304]{Robert A. Wittenmyer} 
\affil{University of Southern Queensland, Centre for Astrophysics, West Street, Toowoomba, QLD 4350 Australia}
\author{Brendan P. Bowler} 
\affil{Department of Astronomy, The University of Texas at Austin, TX 78712, USA}
\author[0000-0002-1160-7970]{Jonathan Horner} 
\affil{University of Southern Queensland, Centre for Astrophysics, West Street, Toowoomba, QLD 4350 Australia}
\author[0000-0002-7084-0529]{Stephen R. Kane} 
\affil{Department of Earth and Planetary Sciences, University of California, Riverside, CA 92521, USA}
\author[0000-0003-0497-2651]{John Kielkopf} 
\affil{Department of Physics and Astronomy, University of Louisville, Louisville, KY 40292, USA}
\author[0000-0002-8864-1667]{Peter Plavchan} 
\affil{George Mason University, 4400 University Drive MS 3F3, Fairfax, VA 22030, USA}

\author[0000-0001-7294-5386]{Duncan J. Wright}
\affil{University of Southern Queensland, Centre for Astrophysics, West Street, Toowoomba, QLD 4350 Australia}
\author[0000-0003-3216-0626]{Brett C. Addison}
\affil{University of Southern Queensland, Centre for Astrophysics, West Street, Toowoomba, QLD 4350 Australia}
\author[0000-0002-7830-6822]{Matthew W. Mengel} 
\affil{University of Southern Queensland, Centre for Astrophysics, West Street, Toowoomba, QLD 4350 Australia}
\author{Jack Okumura}
\affil{University of Southern Queensland, Centre for Astrophysics, West Street, Toowoomba, QLD 4350 Australia}


\author{George Ricker}  
\affiliation{Department of Physics and Kavli Institute for Astrophysics and Space Research, Massachusetts Institute of Technology, Cambridge, MA 02139, USA}

\author{Roland Vanderspek} 
\affiliation{Department of Physics and Kavli Institute for Astrophysics and Space Research, Massachusetts Institute of Technology, Cambridge, MA 02139, USA}

\author[0000-0002-6892-6948]{Sara Seager} 
\affiliation{Department of Physics and Kavli Institute for Astrophysics and Space Research, Massachusetts Institute of Technology, Cambridge, MA 02139, USA}
\affiliation{Department of Earth, Atmospheric and Planetary Sciences, Massachusetts Institute of Technology, Cambridge, MA 02139, USA}
\affiliation{Department of Aeronautics and Astronautics, MIT, 77 Massachusetts Avenue, Cambridge, MA 02139, USA}

\author[0000-0002-4715-9460]{Jon M. Jenkins} 
\affiliation{NASA Ames Research Center, Moffett Field, CA 94035, USA}

\author[0000-0002-4265-047X]{Joshua N.~Winn} 
\affiliation{Department of Astrophysical Sciences, Princeton University, Princeton, NJ 08544, USA}



\author[0000-0002-6939-9211]{Tansu~Daylan}  
\affiliation{Department of Physics and Kavli Institute for Astrophysics and Space Research, Massachusetts Institute of Technology, Cambridge, MA 02139, USA}
\affiliation{Department of Astrophysical Sciences, Princeton University, Peyton Hall, Princeton, NJ 08544}

\author[0000-0002-9113-7162]{Michael~Fausnaugh}
\affiliation{Department of Physics and Kavli Institute for Astrophysics and Space Research, Massachusetts Institute of Technology, Cambridge, MA 02139, USA}

\author[0000-0001-9269-8060]{Michelle Kunimoto} 
\affil{Kavli Institute for Astrophysics and Space Research, Massachusetts Institute of Technology, Cambridge, MA 02139}






\begin{abstract}
The \emph{TESS} mission has enabled discoveries of the brightest transiting planet systems around young stars. These systems are the benchmarks for testing theories of planetary evolution. We report the discovery of a mini-Neptune transiting a bright star in the AB Doradus moving group. \thisstar{} (TOI-4399, TIC 464646604) is a $V_\mathrm{mag}=8.31$ G-dwarf hosting a $3.00_{-0.28}^{+0.32}\,R_\oplus$ mini-Neptune in a 7.7\,day period orbit. \thisstar{} is part of the AB Doradus moving group, one of the youngest and closest associations. Due to its youth, the host star exhibits significant photometric spot modulation, lithium absorption, and X-ray emission. Three 0.06\% transits were observed during Sector-27 of the \emph{TESS} Extended Mission, though these transit signals are dwarfed by the 2\% peak-to-peak photometric variability exhibited by the host star. Follow-up observations with \emph{CHEOPS} confirmed the transit signal and prevented the erosion of the transit ephemeris. \thisstar{} is part of a 50\,AU G-M binary system. We make use of diffraction limited observations spanning 11 years, and astrometric accelerations from \emph{Hipparchos} and \emph{Gaia}, to constrain the orbit of \thisstar{}~B. \thisstar{} is one of the tightest stellar binaries to host an inner planet. As part of a growing sample of bright, young planet systems, \thisstarb{} is ideal for follow-up transit observations, such as those that investigate the evaporative processes driven by high-energy radiation that may sculpt the valleys and deserts in the Neptune population. 
\end{abstract}



\section{Introduction}
\label{sec:introduction}

Young planets offer a time lapse view of the construction of the exoplanet demographics. Planetary systems are thought to undergo rapid evolution within the first hundreds of millions of years after their formation. Follow-up characterization of small young planets helps to test our models for the contraction and mass loss processes that they undergo during this time frame. 

Thousands of close-in Neptunes and super-Earths were discovered by the primary \emph{Kepler} mission \citep[e.g.][]{2013PNAS..11019273P,2015ApJ...809....8B,2018ApJ...860..101Z}. The mechanisms that sculpted the period-radius distribution of these planets can shed light on the early formation and evolution of planetary systems. The evaporation of primordial Hydrogen and Helium envelopes, driven by UV and X-ray radiation from young stars, can reproduce the sub-Neptune desert and radius valley \citep[e.g.][]{2012ApJ...761...59L,2013ApJ...775..105O,2017ApJ...847...29O}. These processes act on rapid timescales because the high-energy fluxes from young stars rapidly decline over time. Mass loss can also be driven by heat leaking out from a planet's deep interior, a process that can last hundreds of millions of years as the planets cool down after accretion \citep[e.g.][]{2013ApJ...776....2L,2018MNRAS.476..759G,2021MNRAS.504.4634G}. Giant impacts within compact super-Earth and Neptune systems may erode the envelopes of some planets, occurring after disk dissipation and before the systems dyanmically cool within the first hundred million years \citep[e.g.][]{2009ApJ...700L.118M,2015MNRAS.448.1751I}. As their primordial gaseous envelopes are stripped away, it is possible that some planets lying in less energetic environments may be replenished by secondary atmospheres \citep[e.g.][]{2020PNAS..11718264K}. Other mechanisms, such as in-situ formation in gas-poor disks, may naturally carve out the current period-radius distribution of small planets \citep{2021ApJ...908...32L,2022arXiv220109898L}, and the radii of small planets may evolve much slower than predicted from run-away mass loss models \citep{2021AJ....161..265D}. 

Young planets can help establish the time-scales for the evolution of the period-radius distribution of the super-Earth and Neptune populations. For some planets, mass loss through photoevaporation occurs throughout their lifetime without significantly changing their radii. Extended atmospheres have been observed for Neptune sized planets about older field stars in X-ray \citep{2012A&A...547A..18E}, Lyman-$\alpha$ \citep{2014ApJ...786..132K,2015Natur.522..459E,2017A&A...605L...7L,2018AA...620A.147B}, and He I \citep{2018Natur.557...68S,2019A&A...623A..58A,2020AJ....159..115K}. However, for planets between 2-4 $R_\oplus$, run-away evaporation may occur. This process strips away the outer primordial envelope of these Neptunes and super-Earths, leaving behind  rocky cores. Real-time measurements of photoevaporation for younger systems can help establish the timescale for this process and its influence on the radius evolution of planets. \citet{2022AJ....163...68Z} detected the Lyman-$\alpha$ transit of the outer planet HD 63433\,c in a 400\,Myr old planetary system. They also reported the lack of a detectable escaping hydrogen atmosphere for the inner planet, suggesting that it may have already undergone run-away evaporation and lost its primordial atmosphere. \citet{2021AJ....162..116R} reported a non-detection of escaping hydrogen for the 650\,Myr old K2-25b, with one possibility for the non-detection being factors related to the star's youth. \citet{2022AJ....163...67Z} reported escaping He I for the 500\,Myr old mini-Neptune HD 73583b, showing an excess  0.68\% absorption for the He I 10830\,\AA{} line. Tentative detections of atmospheric escape have been reported for planets in the V1298 Tau system in Ca II, H-$\alpha$ \citep{2021AJ....162..213F}, and He I \citep{2022MNRAS.509.2969G}

The \emph{K2} mission \citep{2014PASP..126..398H} yielded some of the first young transiting planets \citep[e.g.][]{2016ApJ...818...46M,2016AJ....152...61M,2016Natur.534..658D,2018AJ....156...46V,2018AJ....156..195R,2019AJ....158...79D,2019ApJ...885L..12D,2019MNRAS.484....8L,2019MNRAS.490..698B}, and provided insight into the radius distribution of these planets compared to those about older stars \citep[e.g.][]{2017AJ....154..224R,2021AJ....161..265D}. The \emph{TESS} mission \citep{2015JATIS...1a4003R} is finding young planets that are more suitable for in-depth characterization: DS Tuc A \citep{2019ApJ...880L..17N}, HD\,63433 \citep{2020AJ....160..179M}, and AU Mic \citep{2020Natur.582..497P} are amongst the brightest planet-hosting stars known. These are the best targets for in-depth follow-up studies with the suite of new astronomical observatories coming online this decade. 

In this paper, we present the discovery of a mini-Neptune sized planet transiting the $V=8.3$ star \thisstar{}, a member of the $\sim 120$ Myr old AB Doradus moving group, one of the youngest and closest stellar associations. \thisstar{} was identified as a planet host through a search for planets around active stars \citep{2021AJ....161....2Z}. The star's youth is confirmed from its spectroscopic and photometric characteristics, such as its rapid rotation and significant photometric modulation, strength of lithium absorption, and X-ray emission. The kinematics and independent age estimation of \thisstar{} agree with that of members of the AB Doradus moving group. The shallow 600\,ppm transits of \thisstarb{} were identified in the single sector of observations obtained by \emph{TESS} during the first sector of its ongoing Extended Mission. Subsequent observations via the \emph{CHEOPS} space telescope made it possible to confirm the existence of the transits and improved our ability to predict future transit times for follow-up studies. 

\section{Observations}
\label{sec:obs}

\begin{figure*}
    \centering
    \includegraphics[width=0.8\linewidth]{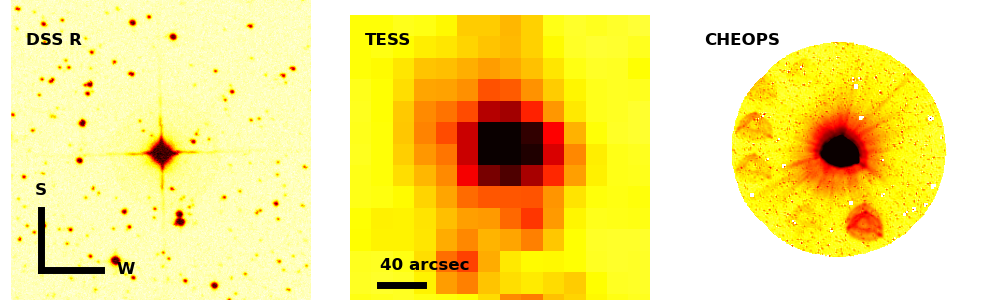}
    \caption{The field around \thisstar{} from \textbf{Left} Deep Sky Survey R' band, \textbf{Middle} \emph{TESS}, and \textbf{Right} \emph{CHEOPS}. The images have approximately the same field sizes (5\arcmin) and orientations.}
    \label{fig:fieldimg}
\end{figure*}

\subsection{Candidate identification with TESS}
\label{sec:TESS}

The Transiting Exoplanet Survey Satellite \citep[\emph{TESS}, ][]{2015JATIS...1a4003R} performed photometric measurements of \thisstar{} in its Sector 27 Camera 2 observations between 2020-07-04 and 2020-07-30. \thisstar{} was observed at two minute cadence via target pixel stamp observations, and was also included as a target of multiple \emph{TESS} Guest Investigator Programs (G03251, D. Huber; G03272, J. Burt). We make use of Simple Aperture Photometry light curves \citep{twicken:PA2010SPIE,morris2020} from the Science Processing Observation Center \citep[SPOC,][]{2016SPIE.9913E..3EJ}, extracted from the two minute target pixel files for subsequent analyses. 

\thisstar{} was identified as a planet candidate by a dedicated search for planets around young stars \citep{2021AJ....161....2Z}. We first identified \thisstar{} as a possible young star via its rotationally modulated light curve. Three transits were identified using a Box-fitting Least Squares search \citep{2002A&A...391..369K} on the light curve after detrending using a high order spline-fitting procedure \citep{2014PASP..126..948V}. \thisstar{} was also identified as a threshold crossing event in both the SPOC pipeline \citep{2016SPIE.9913E..3EJ} and the MIT QuickLook Pipeline \citep{2020RNAAS...4..204H}. The SPOC threshold crossing event diagnostic tests indicate the transit is on target to within $5.3 \pm 2.5\arcsec$. However, it did not survive the TOI vetting process initially because of the large amplitude of the residuals in the detrended light curves, which is due to the intrinsic variability of the star. The candidate was promoted to become a \emph{TESS} Object of Interest (TOI) after the confirmation of the photometric transit signals using the \emph{CHEOPS} observations (Section~\ref{sec:CHEOPS}). 

To account for the instrumental noise due to spacecraft motion and stellar variability in the light curves, we perform a simultaneous detrending using the spacecraft quarternions and basis splines following a similar process to that described in \citet{2019ApJ...881L..19V}. We iteratively fit for linear coefficients of the mean, standard deviation, and skew terms of the three quarternions, together with a spline matrix created by the \textsc{lightkurve} package \citep{2018ascl.soft12013L}. The detrended light curve is adopted for our subsequent analyses in Section~\ref{sec:analysis}.    

\begin{figure*}
    \centering
    \includegraphics[width=0.8\linewidth]{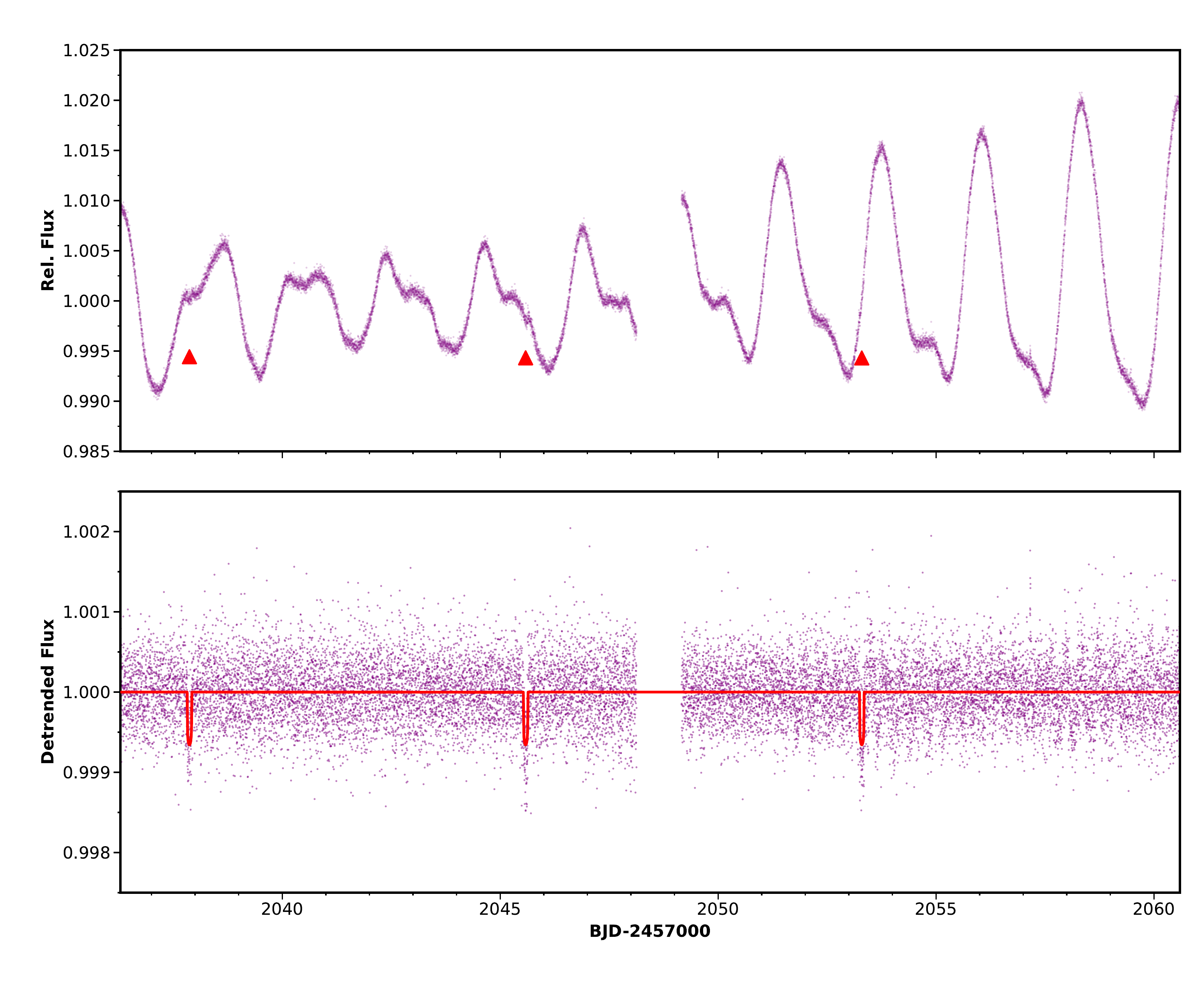}
    \caption{Sector 27 2\,min cadence \emph{TESS} photometry of the \thisstar{} system. \textbf{Top:} Simple Aperture \emph{TESS} light curve of \thisstar{}. The three detected transits of \thisstarb{} are indicated by the arrow marks. \textbf{Bottom:} The custom detrended \emph{TESS} light curve of \thisstar{} are shown, with the best fit model overlaid in red.}
    \label{fig:lightcurve}
\end{figure*}

\begin{figure*}
    \centering
    \includegraphics[width=0.8\linewidth]{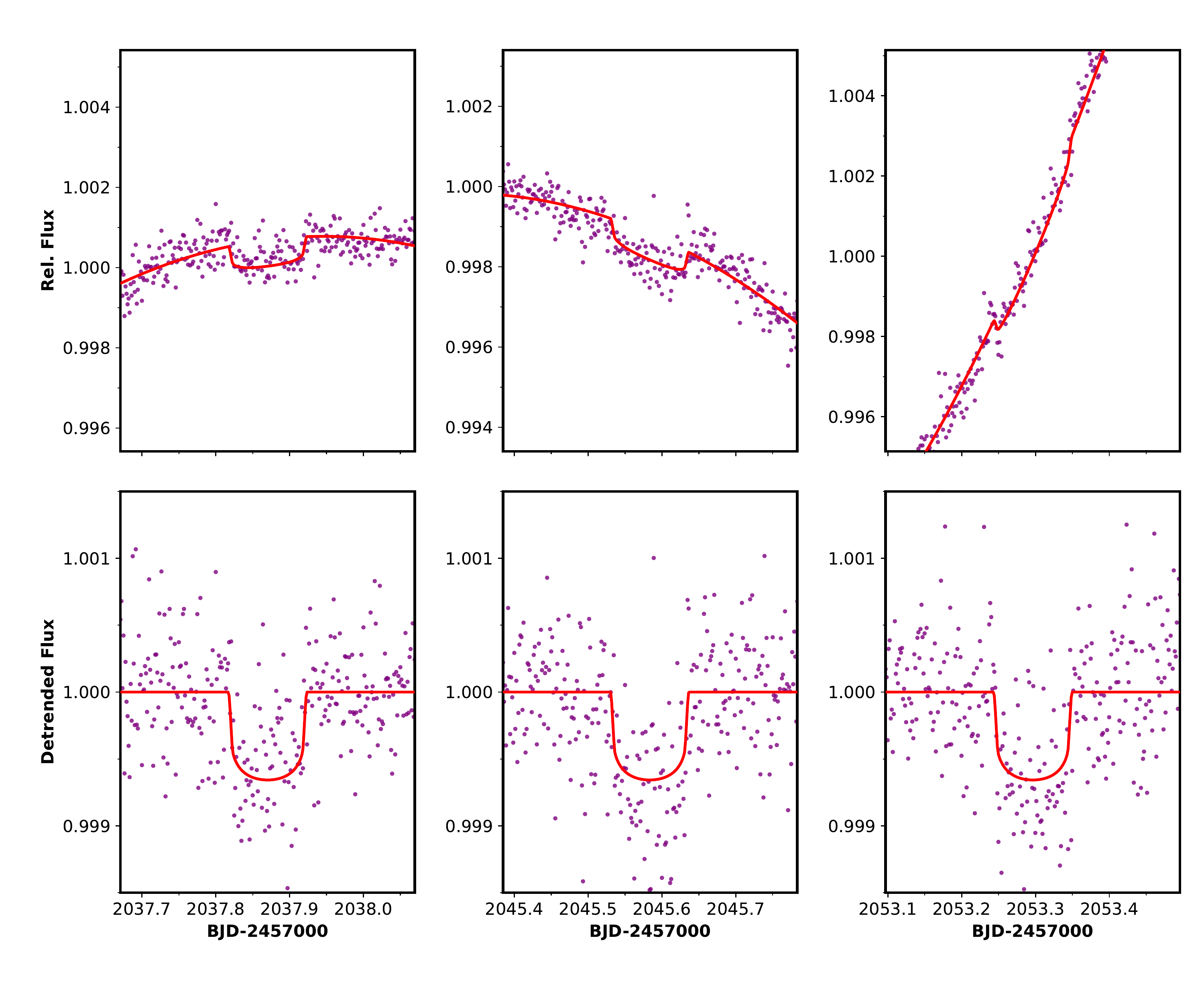}
    \caption{Individual transits of \thisstarb{} from \emph{TESS} Sector-27. The \textbf{top} row shows the transits pre-detrending, the \textbf{bottom} row shows the same transits post-detrending. The best fit transit model is overlaid in red.}
    \label{fig:individual_lightcurve}
\end{figure*}

\begin{figure}
    \centering
    \includegraphics[width=\linewidth]{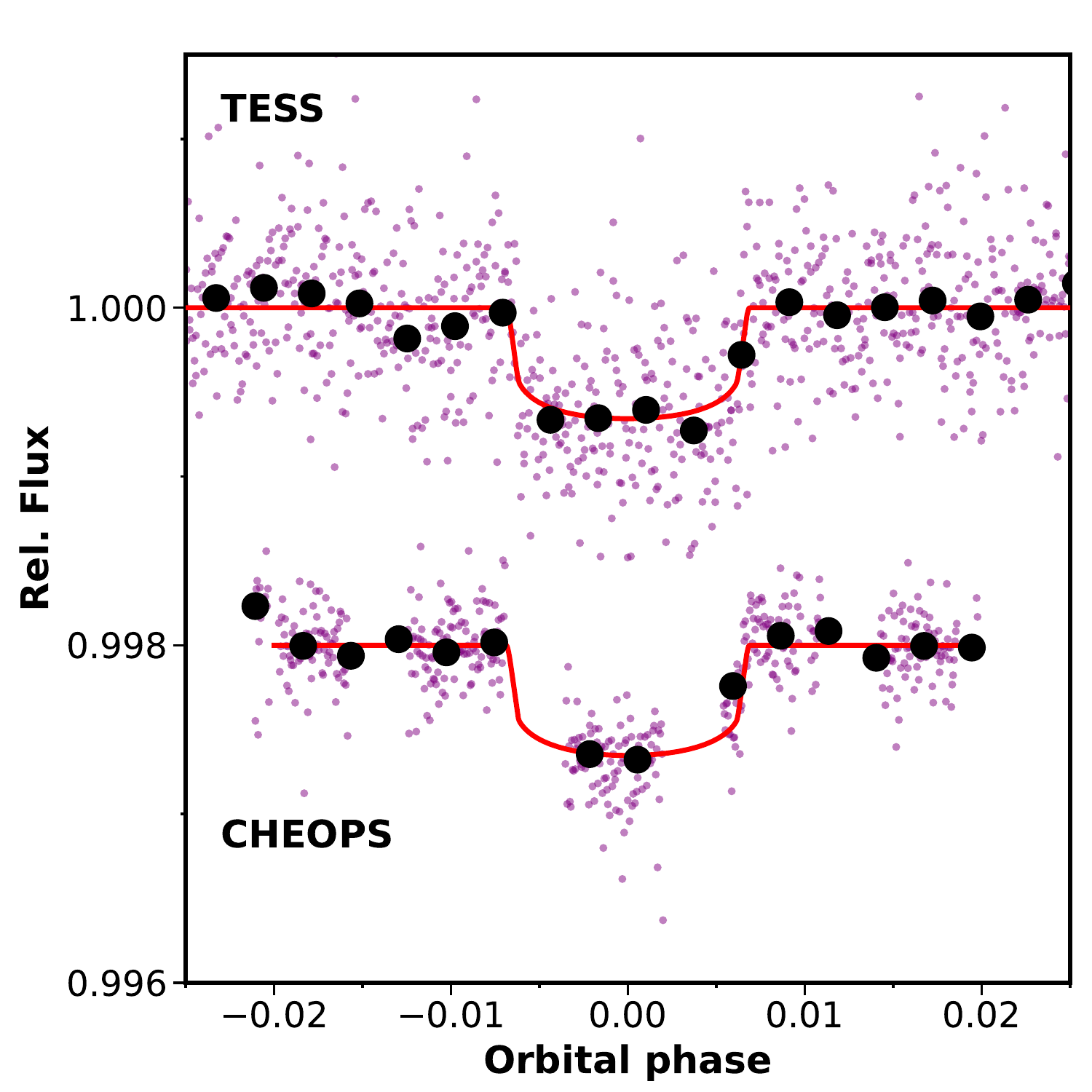}
    \caption{Phase folded transit light curves from \emph{TESS} \textbf{(Top)} and \emph{CHEOPS} \textbf{(Bottom)}. The black points show the dataset binned at 30\,minute intervals. The red lines show the respective models from our global analysis. }
    \label{fig:lcbinned}
\end{figure}

The corrected \emph{TESS} light curve from Sector-27 is shown in Figure \ref{fig:lightcurve}. Figure~\ref{fig:individual_lightcurve} shows close-ups of individual transits of \thisstarb{} in the raw and detrended \emph{TESS} light curves. Figure~\ref{fig:lcbinned} shows the phase-folded \emph{TESS} transit light curve and its best fit model from our global modeling (Section~\ref{sec:analysis}). 

\subsection{Follow-up photometry with \emph{CHEOPS}}
\label{sec:CHEOPS}
To confirm the existence and constrain the transit ephemeris of \thisstarb{}, we obtained space-based photometric observations using the ESA's CHaracterising ExOPlanets Satellite \citep[\emph{CHEOPS}][]{2021ExA....51..109B}, through the \emph{CHEOPS} Guest Observers Programme (AO2-005). \emph{CHEOPS} is a 0.32\,m Ritchey-Chr\'{e}tien telescope in a nadir-locked  700\,km low-Earth orbit, along the Earth's day/night terminator, with a period of 98.725 minutes. The telescope has a field of view of $17\arcmin \times 17 \arcmin$, and is instrumentally defocused with a point-spread function of $16\arcsec$ at a plate scale of $1\arcsec \mathrm{pixel}^{-1}$. 

A transit of \thisstarb{} was observed by \emph{CHEOPS} between 2021-08-19 23:59 to 2021-08-20 07:33 UTC (visit ID 1568350). The visit consists of five orbits, with an exposure time of 17\,s coadded onboard to a cadence of 34\,s. The visit had an observing efficiency of 61\% (fraction of time on target), with interruptions primarily due to Earth stray light. A total of 498 exposures were obtained, of which, nine exposures were affected by stray light and Earth occultation, eight by South Atlantic Anomaly crossings. The observations were reduced by the \emph{CHEOPS} Data Reduction Pipeline v13.1.0 \citep{2020A&A...635A..24H}, accounting for bias, dark current, flat fielding, bad pixel correction, smear contamination and linearization. We adopt the optimal aperture light curve, with a circular aperture of 31 pixels, for our analysis. Aperture contamination by background stars is estimated to be at the $6\times10^{-4}$ level, and is accounted for via the simulations from the data reduction pipeline.  

\begin{figure}
    \centering
    \includegraphics[width=\linewidth]{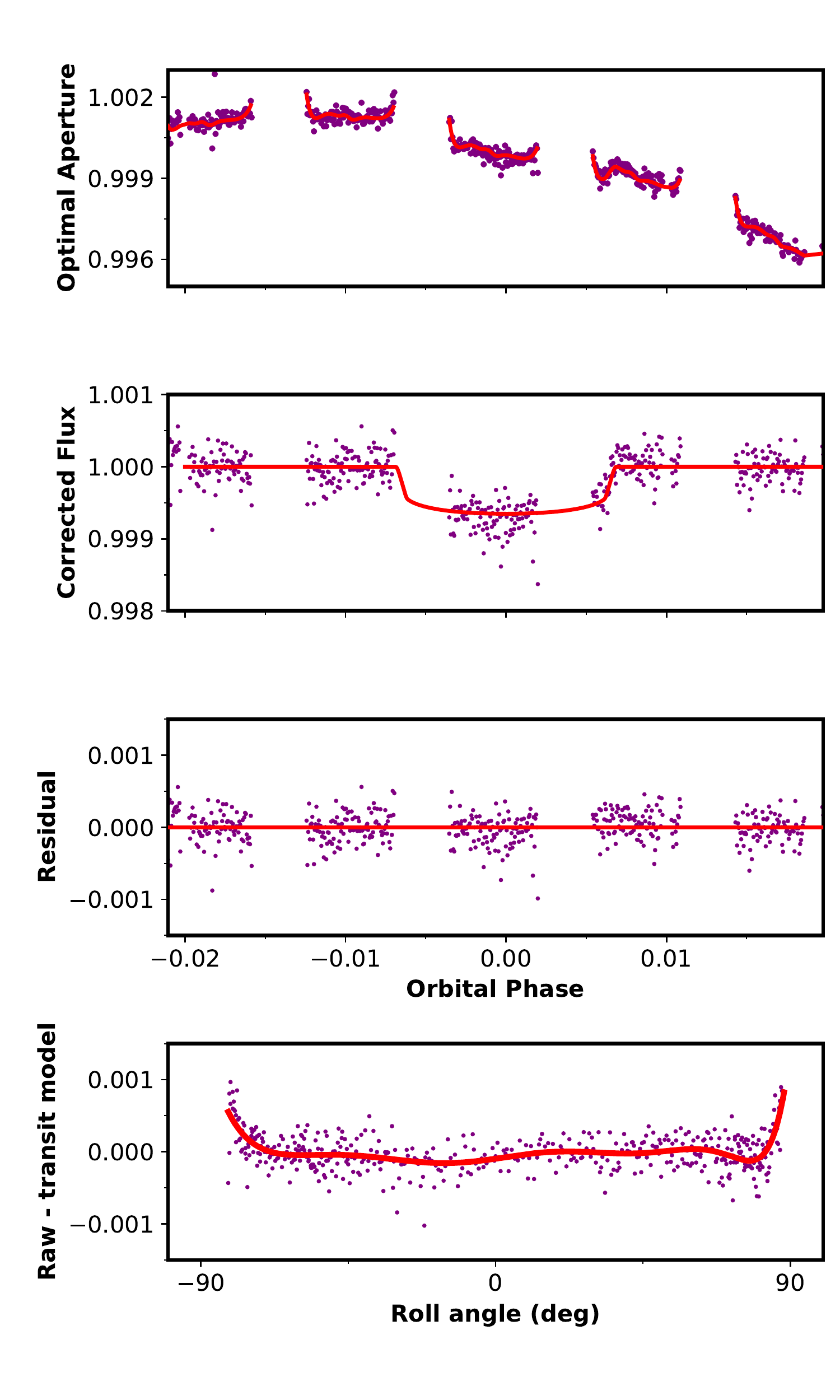}
    \caption{Transit of \thisstarb{} as captured by \emph{CHEOPS} on 2021-08-19 over five \emph{CHEOPS} orbits. The \textbf{first panel} panel shows the pre-detrending light curve extracted at an optimal aperture of 31 pixels. The large scale smooth variation is induced by the spot modulation of the target star, and is accounted for in our model via a fourth order polynomial. Orbital-timescale variations are also present, and are accounted for via a 5th order polynomial fit between the spacecraft roll angle and the light curve residuals, after subtraction of the transit model and the spot modulation trend. The \textbf{second} panel shows the detrended \emph{CHEOPS} light curve. The \textbf{third} panel shows the residuals after subtraction of the transit, spot modulation signal, and instrumental systematics models. The \textbf{fourth} panel shows the correlation between the light curve, after removal of the transit and spot modulation signals, and the spacecraft roll angle, with the best fit model overlaid. }
    \label{fig:cheopsmodel}
\end{figure}

The \emph{CHEOPS} light curve exhibits a smooth hours-long trend that can be attributed to the spot modulated rotational variations of \thisstar{}. Shorter timescale instrumental variations associated with the spacecraft roll angle are seen on the orbital timescales \citep{2021arXiv211108828M}. We model the spot modulation over the five orbits of observations with a 4th order polynomial after removal of the transit model. We also fit for a 5th order correlation between the spacecraft roll angle and the resulting light curve residuals. This modeling is performed simultaneous to the global fit of the \emph{TESS} light curve and associated parameters, such that the uncertainties from this detrending process are fully accounted for in our results (Section~\ref{sec:analysis}). The raw and corrected \emph{CHEOPS} light curves are shown in Figure~\ref{fig:cheopsmodel}, and the binned and corrected phase-folded data is compared with that from \emph{TESS} in Figure~\ref{fig:lcbinned}. 

\begin{figure}
    \centering
    \includegraphics[width=0.9\linewidth]{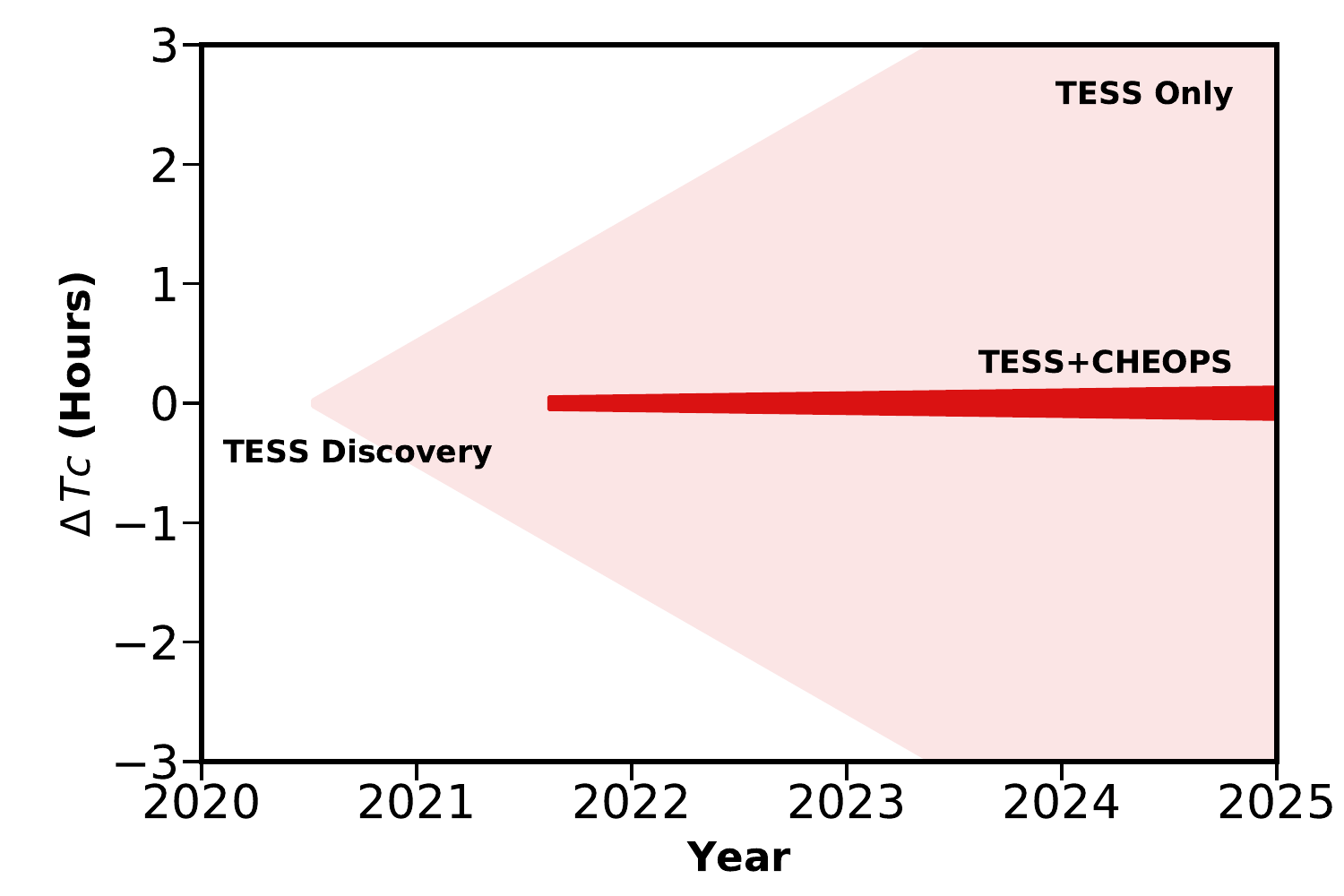}
    \caption{Transit timing uncertainty for \thisstarb{} with and without follow-up \emph{CHEOPS} observations. With only the single sector of \emph{TESS} observations, the timing uncertainty erodes by $\sim 1\,\mathrm{hr\,yr}^{-1}$, quickly making targeted transit follow-up difficult. The \emph{CHEOPS} transit allowed the transit ephemeris to be preserved. }
    \label{fig:terrpropagation}
\end{figure}

The \emph{CHEOPS} observation was most crucial in refining the ephemeris of \thisstarb{}. With a transit depth of 600\,ppm, detecting the photometric transit event with ground-based facilities is difficult. With only three transits available during the single sector of \emph{TESS} observations over the entire primary and first extended mission, the \emph{TESS} ephemeris would have quickly become stale. The ephemeris uncertainty using \emph{TESS} observations alone would have been 5.2 hours after five years, making any transit follow-up studies more difficult to schedule. The single \emph{CHEOPS} transit reduced the five-year transit timing uncertainty to 8 minutes. Figure~\ref{fig:terrpropagation} illustrates the reduction in transit ephemeris uncertainty enabled by the \emph{CHEOPS} observation. 

\subsection{Clearing of nearby eclipsing binaries from Las Cumbres Observatory}
\label{sec:lco}

The large \emph{TESS} and \emph{CHEOPS} point-spread functions allow for multiple stars within a photometric aperture to be the source of a photometric transit detection. We obtained a transit observation of \thisstarb{} with the Las Cumbres Observatory Global Network \citep[LCOGT][]{2013PASP..125.1031B} 1\,m telescope at Las Campanas Observatory on 2021-05-26 UTC. The observations were performed in the $g'$ band, and were defocused to a full width at half maximum of $7\arcsec$ due to the brightness of the target star. The observations showed that no nearby stars visually separable with \thisstar{} within $1\arcmin$ exhibited eclipsing or transiting events during the predicted transit of \thisstarb{}. Though the transit of \thisstarb{} was not detected due to its shallow depth, this LCOGT observation cleared nearby stars of being stellar eclipsing binaries. 

\subsection{Reconnaissance Spectroscopy}
\label{sec:reconspec}

To characterize the target star and confirm its youth spectroscopically, we obtained three observations of \thisstar{} with the High Resolution Spectrograph \citep[HRS][]{2014SPIE.9147E..6TC} on the Southern African Large Telescope \citep[SALT;][]{2006SPIE.6267E..0ZB} in October 2020. HRS is a fibre fed echelle spectrograph with a resolving power of $R \sim 65,000$ over the wavelength range of $3700-8900\,\AA$. Spectral extraction was performed via the MIDAS pipeline \citep{KniazevMN482016, KniazevSALT2017}\footnote{\href{http://www.saao.ac.za/~akniazev/pub/HRS_MIDAS/HRS_pipeline.pdf}{http://www.saao.ac.za/\texttildelow akniazev/pub/HRS\_MIDAS/HRS\_pipeline.pdf}}. These observations confirmed the presence of the lithium 6708\,\AA{} doublet and strong chromospheric emission in the Calcium H \& K line cores. The lack of significant radial velocity variations or secondary sets of spectral lines indicated the transiting candidate was not an obvious blended eclipsing stellar binary system.

We obtained 16 observations of \thisstar{} using the CHIRON fiber fed cross-dispersed echelle spectrometer at the SMARTS 1.5-meter telescope located at Cerro Tololo Inter-American Observatory, Chile \citep{2013PASP..125.1336T}. CHIRON has a spectral resolving power of $R=80,000$ over the wavelength range of 4100 to $8700$\,\AA{}. Spectra from CHIRON were reduced as per \citet{2021AJ....162..176P}. Radial velocities were measured following the procedure from \citet{2021AJ....161....2Z} via a least-squares deconvolution of each observation against a synthetic non-rotating template generated from the ATLAS9 model atmospheres \citep{Castelli:2004}. The radial velocity measurements are presented in Table \ref{tab:RVs}. 

In addition, we follow \citet{2021AJ....161....2Z} and measure the spectroscopic atmospheric parameters of \thisstar{} from the CHIRON spectra. We match the spectrum of \thisstar{} against an interpolated library of $\sim 10,000$ observed spectra pre-classified by the Spectroscopic Classification Pipeline \citep{2012Natur.486..375B}, finding that \thisstar{} has an effective temperature of $5991\pm50$\,K, surface gravity of $4.46\pm0.02$\,dex, and metallicity of $\mathrm{[M/H]}=-0.05\pm0.10$\,dex. We adopt the spectroscopic effective temperature as a prior in the global modeling described in Section~\ref{sec:analysis}. 

We also used the MINERVA-Australis telescope array for further reconnaissance of \thisstar{}. MINERVA-Australis is an array of four identical 0.7\,m CDK700 telescopes located at Mt Kent Observatory, Australia \citep{2019PASP..131k5003A}. The telescopes feed into a single KiwiSpec echelle spectrograph with a spectral resolving power of $R\approx80,000$ over the wavelength region of $4800-6200$\,\AA. The instrument is environmentally controlled inside a vacuum chamber, and simultaneous wavelength calibration is provided by two calibration fibers that bracket the science fibers on the detector, each fed from a quartz lamp via an iodine cell. Radial velocities are measured from the extracted spectra as per our CHIRON analyses described above, via a least-squares deconvolution between the observations and the synthetic non-rotating template. The radial velocities are provided in Table~\ref{tab:RVs}. 

\begin{deluxetable}{rrrr}
\tablewidth{0pc}
\tabletypesize{\scriptsize}
\tablecaption{
    Radial Velocity Measurements of \thisstar{}
    \label{tab:RVs}
}
\tablehead{ \\
    \multicolumn{1}{c}{BJD}   &
    \multicolumn{1}{c}{RV} &
    \multicolumn{1}{c}{RV Error} &
    \multicolumn{1}{c}{Instrument} \\
    \multicolumn{1}{c}{}   &
    \multicolumn{1}{c}{(km s$^{-1}$)} &
    \multicolumn{1}{c}{(km s$^{-1}$)} &
    \multicolumn{1}{c}{}
}
\startdata
2459305.91348 & 8.131 & 0.100 & CHIRON \\
2459309.87630 & 8.211 & 0.134 & CHIRON \\
2459311.89601 & 8.220 & 0.073 & CHIRON \\
2459320.89052 & 8.152 & 0.082 & CHIRON \\
2459327.86151 & 8.246 & 0.079 & CHIRON \\
2459335.86210 & 8.340 & 0.120 & CHIRON \\
2459337.89892 & 8.123 & 0.099 & CHIRON \\
2459339.86391 & 8.144 & 0.087 & CHIRON \\
2459459.55982 & 8.069 & 0.087 & CHIRON \\
2459461.57939 & 8.255 & 0.083 & CHIRON \\
2459463.56573 & 8.285 & 0.072 & CHIRON \\
2459464.59690 & 8.273 & 0.068 & CHIRON \\
2459465.56939 & 8.271 & 0.282 & CHIRON \\
2459467.53083 & 8.394 & 0.092 & CHIRON \\
2459478.48380 & 8.459 & 0.098 & CHIRON \\
2459479.48429 & 8.123 & 0.119 & CHIRON \\
2459480.51450 & 8.355 & 0.093 & CHIRON \\
2459481.51666 & 8.466 & 0.096 & CHIRON \\
2459482.55682 & 8.350 & 0.068 & CHIRON \\
2459485.52901 & 8.485 & 0.062 & CHIRON \\
2459506.52501 & 8.237 & 0.071 & CHIRON \\
2459508.52808 & 8.107 & 0.102 & CHIRON \\
2459319.14408 & 8.841 & 0.104 & MINERVA-Australis Tel 3 \\
2459324.21159 & 8.964 & 0.175 & MINERVA-Australis Tel 3 \\
2459332.10157 & 8.914 & 0.178 & MINERVA-Australis Tel 3 \\
2459348.10572 & 9.096 & 0.165 & MINERVA-Australis Tel 3 \\
2459452.00282 & 9.152 & 0.143 & MINERVA-Australis Tel 3 \\
2459453.04237 & 8.976 & 0.150 & MINERVA-Australis Tel 3 \\
2459453.99779 & 9.120 & 0.140 & MINERVA-Australis Tel 3 \\
2459456.02452 & 9.171 & 0.139 & MINERVA-Australis Tel 3 \\
2459476.98480 & 8.961 & 0.109 & MINERVA-Australis Tel 3 \\
2459319.14408 & 9.208 & 0.161 & MINERVA-Australis Tel 4 \\
2459324.21159 & 8.766 & 0.101 & MINERVA-Australis Tel 4 \\
2459348.10572 & 8.995 & 0.167 & MINERVA-Australis Tel 4 \\
2459452.00282 & 9.177 & 0.194 & MINERVA-Australis Tel 4 \\
2459453.04237 & 9.017 & 0.146 & MINERVA-Australis Tel 4 \\
2459453.99779 & 9.265 & 0.148 & MINERVA-Australis Tel 4 \\
2459476.98480 & 9.121 & 0.179 & MINERVA-Australis Tel 4 \\
2459319.14408 & 8.955 & 0.176 & MINERVA-Australis Tel 5 \\
2459324.21159 & 9.085 & 0.137 & MINERVA-Australis Tel 5 \\
2459332.10157 & 8.741 & 0.123 & MINERVA-Australis Tel 5 \\
2459348.10572 & 9.077 & 0.181 & MINERVA-Australis Tel 5 \\
\enddata
\end{deluxetable}

\begin{figure*}
    \centering
    \includegraphics[width = 0.7\linewidth]{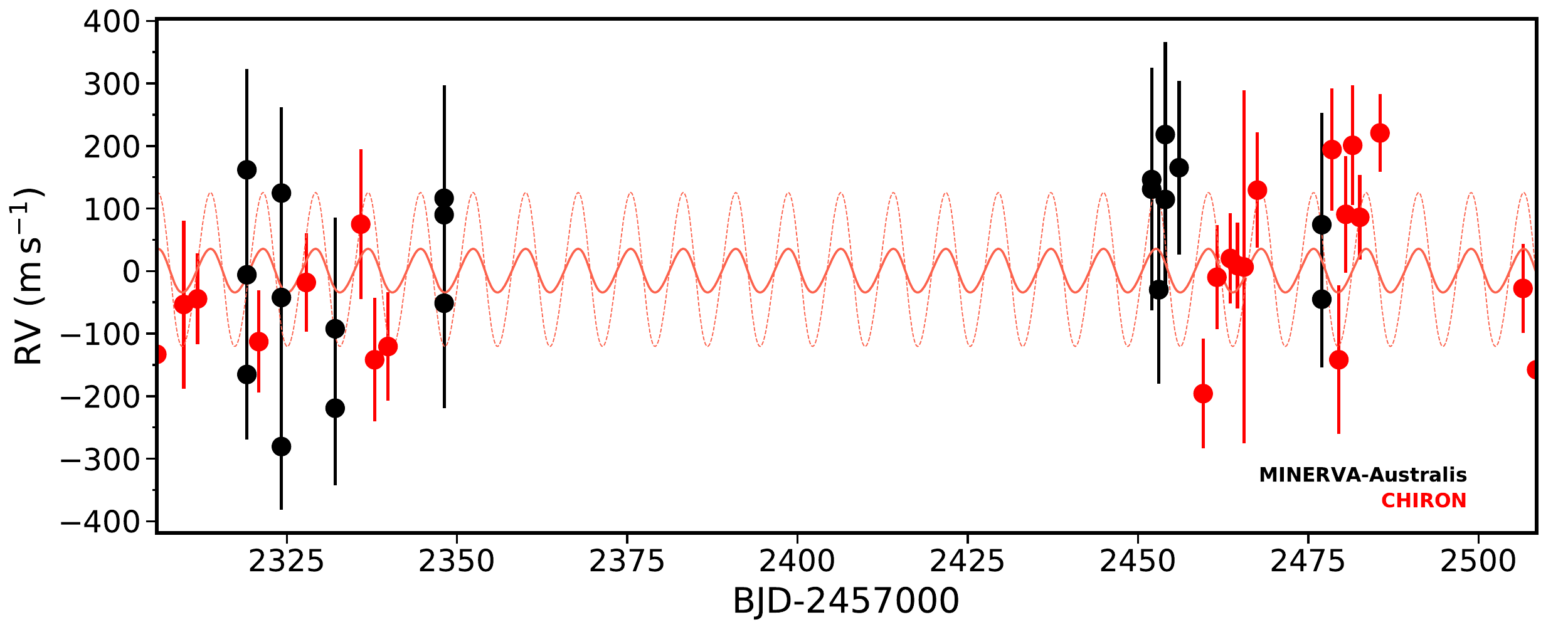}
    \caption{Radial velocities for \thisstar{} from CHIRON (red) and MINERVA-Australis (black), with errorbars representing the quadrature addition of the observational uncertainties and the best fit jitter. No orbital variations were detected, as expected for a small planet about an active, rapidly rotating star. The velocity orbit upper limits at $1\sigma=0.35\,M_\mathrm{jup}$ and $3\sigma=1.27\,M_\mathrm{jup}$ are shown in the solid and dashed red curves respectively.}
    \label{fig:rvs}
\end{figure*}

\section{Age and membership}
\label{sec:age}

\thisstar{} shares space velocities with the AB Doradus moving group (Section~\ref{sec:abdor}). The age of the group has been estimated to be between 50-150 Myr. The group has also been linked to be co-evolving with the Pleiades. We adopt an AB Doradus age of 120 Myr for the remainder of this discussion.  In addition, a neighborhood search for tangentially co-moving stars with \thisstar{} reveals a tentative population of stars, with an age of $\sim 120$\,Myr, determined by their rotation periods. The following sections examine the kinematic properties of \thisstar{} and the association, as well as the photometric and spectroscopic characteristics of \thisstar{} that identify its youth independent of the kinematics.

As accretion dwindles, young stars spin up as they conserve their angular momentum while they contract in radius. The spin up leads to increased magnetic activity, large photospheric spots, and increased chromospheric activity. Both rotation and activity then decay with time as angular momentum is lost through the stellar wind over the main sequence lifetime of the star. The \emph{TESS} light curve of \thisstar{} exhibits significant rotational spot modulation, which first drew our attention to its potential youth. The presence of X-ray emission and lithium absorption for the Sun-like host star confirm its youth. A flare event was also observed during the second orbit of the \emph{TESS} sector, consistent with behavior expected for a young star. 

\subsection{Kinematics}
\label{sec:kinematics}

\subsubsection{The AB Doradus moving group}
\label{sec:abdor}

Accurate estimation of the age of a single star is notoriously difficult \citep{2010ARA&A..48..581S}. Stars in clusters and associations have accurate age estimates since the population can be assessed as a whole, the stars can be seen to be co-evolving based on their color-magnitude, rotation, and lithium abundance distributions. Planets found in co-evolving populations offer much more stringent tests on the temporal evolution of planet properties. 

\citet{2004ApJ...613L..65Z} identified a set of stars within $\sim 50$\,pc co-moving with AB Doradus. The star AB Doradus itself is amongst the closest (15 pc) and most well studied young stars. Membership of the group has been revised based on chemo-kinematic analyses of the homogeneity of the stars \citep{2009A&A...508..833D} and updated kinematics from new missions \citep[e.g.][]{2013ApJ...762...88M, 2018ApJ...856...23G}. Today, dozens of bona-fide members define the extent and characteristics of the group, and its age has been estimated to range from 50\,Myr \citep{2004ApJ...613L..65Z}, $\sim 100-120$\, Myr \citep{2005ApJ...628L..69L}, $>110$ Myr \citep{2013ApJ...766....6B}, to $\sim 150$ Myr \citep{2015MNRAS.454..593B}.

\begin{figure*}
    \centering
    \includegraphics[width=0.8\linewidth]{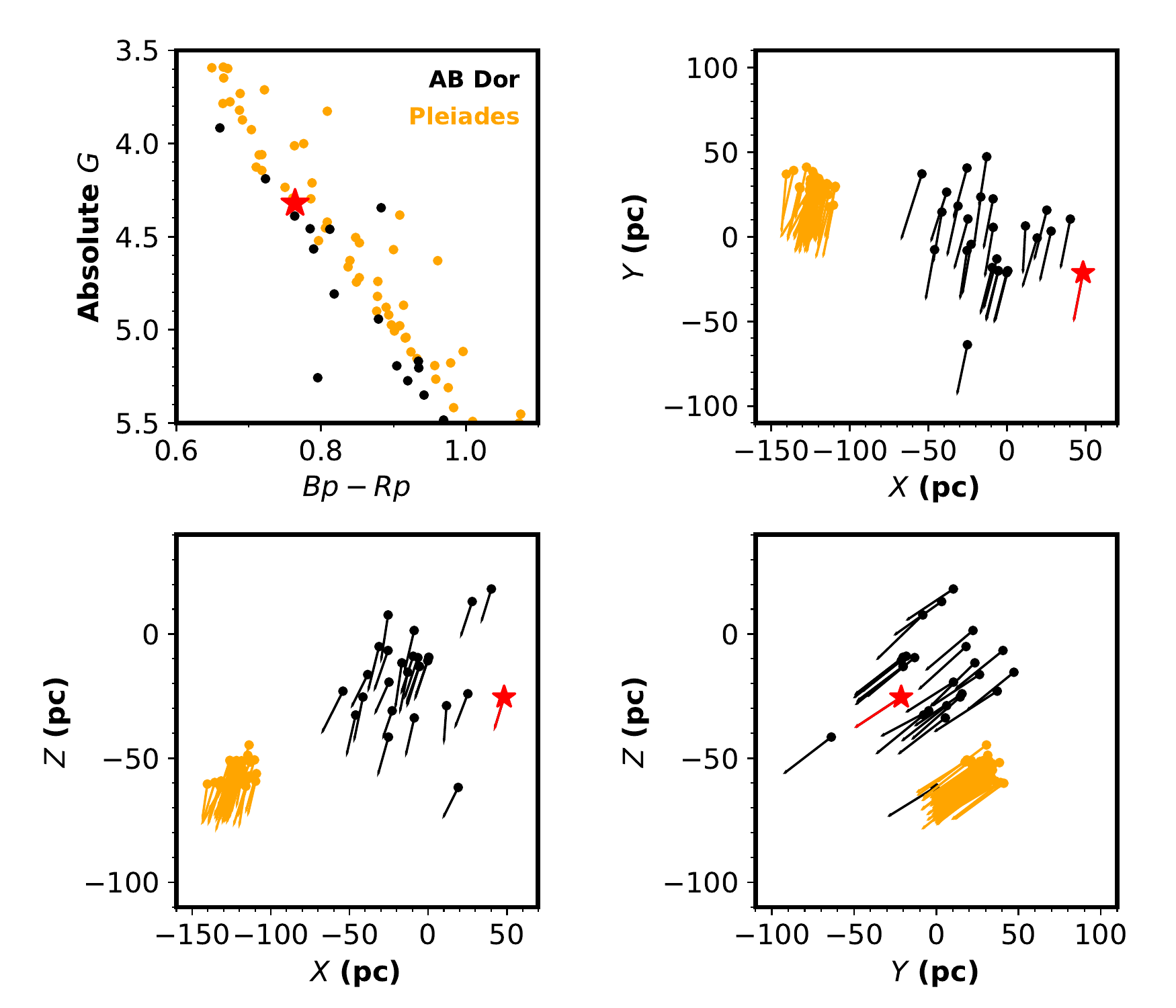}
    \caption{\thisstar{} can be placed kinematically in the $\sim 120$ Myr old AB Doradus moving group. The figure shows the photometric and space motion properties of stars in AB Doradus and Pleiades from \citet{2018ApJ...856...23G}. Both groups have ages of $\sim 120$ Myr, and are thought to be co-moving \citep{2005ApJ...628L..69L,2007MNRAS.377..441O,2019AJ....158..122K}. The \textbf{top left} panel shows the \emph{Gaia} color magnitude diagram for the two clusters. The remaining panels show the galactic positions $X,Y,Z$ and motions $u,v,w$ of the selection. \thisstar{} is marked by the red star in all panels. The local standard of rest has not been corrected for in the velocities. }
    \label{fig:uvw}
\end{figure*}

Figure~\ref{fig:uvw} shows the color-magnitude and space-motion of \thisstar{} alongside the AB Doradus moving group and the Pleiades cluster. \thisstar{} shares kinematic properties with the moving group, and has been classified as a bona-fide member via \emph{Hipparcos} \citep{2013ApJ...762...88M} and \emph{Gaia} \citep{2018ApJ...856...23G,2020AJ....159..166U} space-motions. 

The group has long been linked to Pleiades cluster due to their shared kinematic velocities \citep{2005ApJ...628L..69L,2007MNRAS.377..441O}. Recent mapping of new low density moving groups from Gaia \citep{2019AJ....158..122K} suggest that the AB Doradus moving group, alongside newly identified Theia 301 and Theia 369 associations, form a long tidal tail streaming away from the Pleiades \citep{2021ApJ...915L..29G}. 

Though more distant (130 pc), the Pleiades cluster contains $\sim 1000$ members, and its age has been more thoroughly investigated than AB Doradus. Estimates converge to $\sim 120$ Myr via non-rotating isochrones \citep[e.g.][]{1993A&AS...98..477M}, lithium depletion boundary \citep[e.g.][]{1998ApJ...499L.199S,2004ApJ...614..386B}, and 3D rotational isochrones \citep[e.g.][]{2015ApJ...807...58B}. The spectroscopic and photometric characteristics of \thisstar{} agree with members of the Pleiades, reaffirming our adopted age of $\sim 120$ Myr for the system. 

\subsubsection{An independent rotation sequence for co-moving and co-evolving stars}
\label{sec:comove}

The sparsity and spread of the AB Doradus moving group makes it difficult to securely identify true members of the group. The canonical membership list has been evolving with each new chemo-kinematic dataset since it was originally defined in \citet{2004ApJ...613L..65Z}. Upcoming \emph{Gaia} releases with refined radial velocities for fainter distributions of cool stars may help redefine groups like AB Doradus, and its links with other associations. 

To ensure that our age estimation does not hinge on literature classifications of \thisstar{}, we also independently search for a co-moving population that may confirm its youth. 

\begin{figure*}
    \centering
    \includegraphics[width=0.8\linewidth]{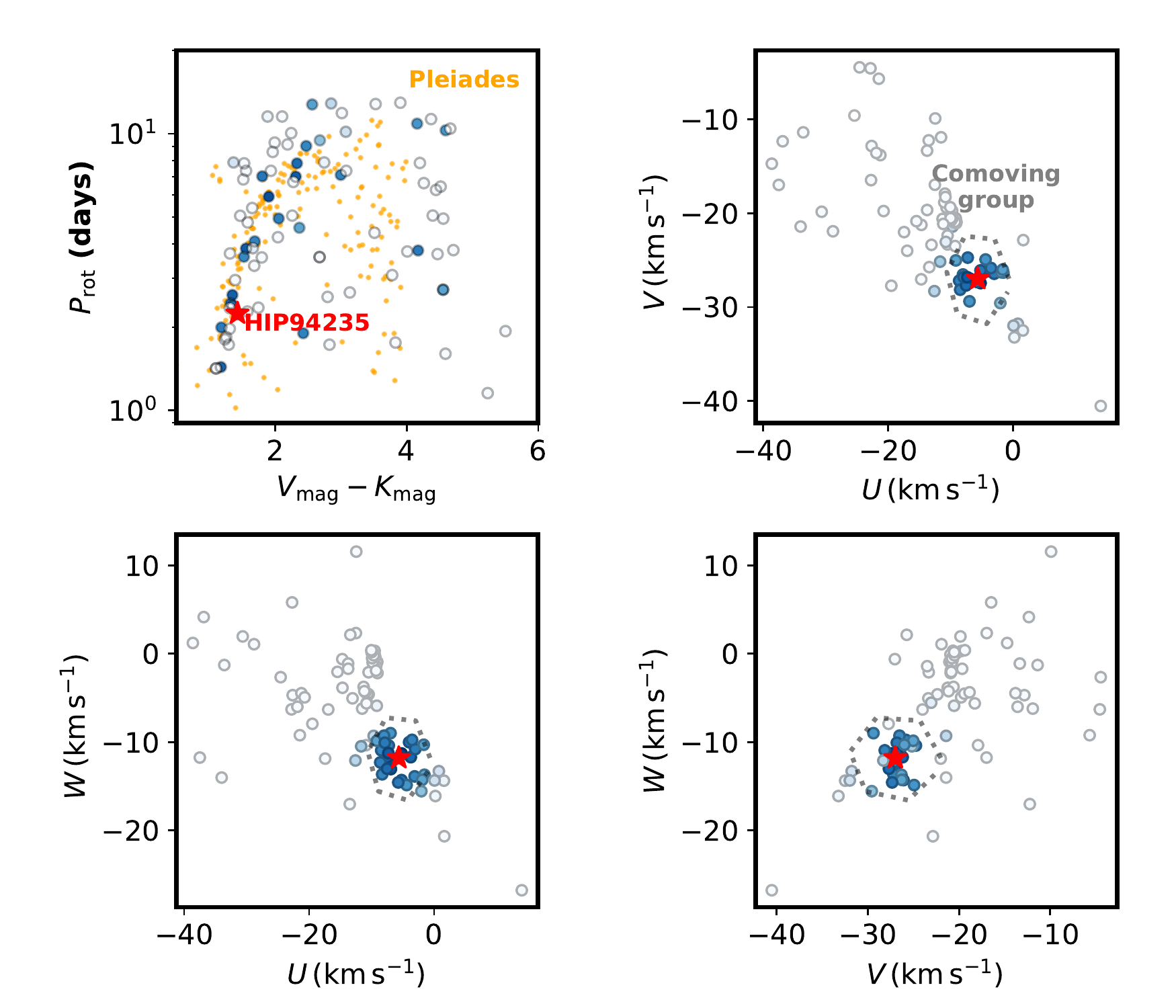}
    \caption{The distribution of stars sharing tangential velocities of \thisstar{} identified in our neighborhood search. \textbf{Top left} We find 140 stars to exhibit clear rotational signatures in their \emph{TESS} light curves. The blue points are located within 5\,\kms{} of \thisstar{} in $U$, $V$, and $W$. These blue points form rotation sequence that agrees with the expected age of \thisstar{} at 120\,Myr. The Pleiades sequence from \citet{2016AJ....152..113R} is shown in yellow. The \textbf{remaining panels} show the $U$, $V$, and $W$ velocity distribution of the tangentially co-moving stars. A clear group can be found around \thisstar{}, and is marked out by the 5\,\kms{} boundary in grey. }
    \label{fig:uvw_friends}
\end{figure*}

We follow \citet{2021arXiv210206066T} and search for stars within 50\,pc of \thisstar{} that share its tangential velocity to within $5\,\mathrm{km\,s}^{-1}$ using the \textsc{comove} package\footnote{\url{https://github.com/adamkraus/Comove}}. In this search, we assume the radial velocity of \thisstar{} for all neighborhood stars.  The search returns a set of $\sim 2000$ stars between $3 < G < 20$ in magnitude. We then query for \emph{TESS} MIT QLP light curves, returning 300 matches up to $T_\mathrm{mag} = 13.5$. We apply a Lomb Scargle period search for each light curve, with an upper limit on the rotation period of $P_\mathrm{rot} < 13$. After manual examination, we find 140 stars that exhibit rotational modulation in their light curves with secure periods. 

Figure~\ref{fig:uvw_friends} shows the rotation distribution of these 140 stars relative to the Pleiades. The vast majority of these stars are relatively bright, and have well measured radial velocities from \emph{Gaia}. We find 21 stars with space motion velocities within 5\,\kms of \thisstar{}, these are coloured in blue for clarity. This population forms a rotation sequence that agrees with \thisstar{}. This set of co-evolving stars are presented in Table~\ref{tab:comove}. 

\begin{figure}
    \centering
    \includegraphics[width=0.9\linewidth]{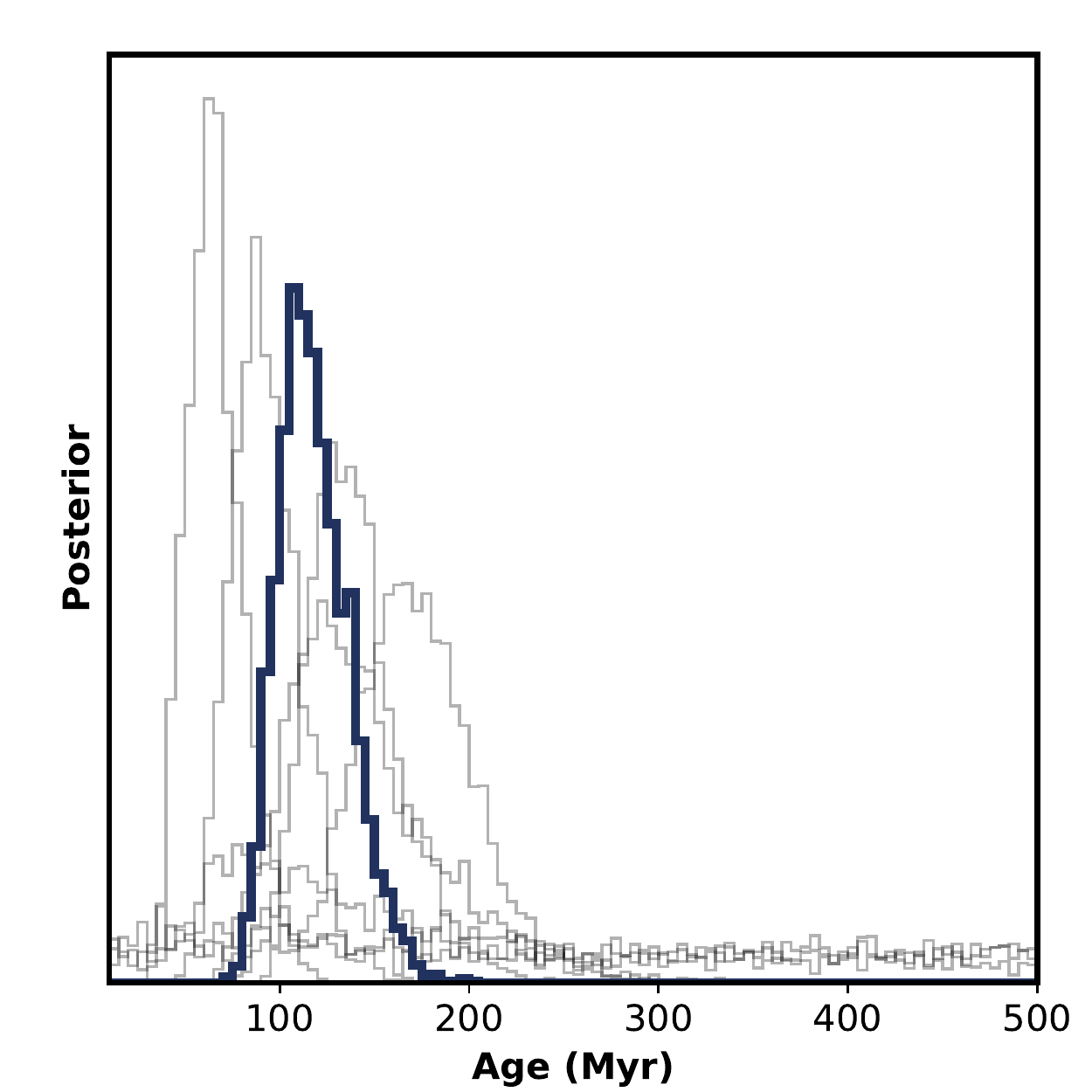}
    \caption{The age posterior of a subsample of F,G,K stars that are co-moving with \thisstar{}. We fit the age relationship from \citet{2008ApJ...687.1264M} to this population. The age posterior of each individual star is shown in grey, and the joint posterior in blue. We find a best-fit age of $118_{-16}^{+18}$ Myr to the distribution.}
    \label{fig:ageposterior}
\end{figure}

To estimate the age of the distribution, we fit its color and rotation period with a rotation-age relationship. We select eight F, G, K stars ($0.1 < B-V < 0.8$) within our sample that lie on the slow sequence of the rotation sequence, and model their rotation periods with the age-color-rotation relationship from \citet{2008ApJ...687.1264M}. The posterior age distribution for these eight stars, and their joint posterior, is shown in Figure~\ref{fig:ageposterior}. We find that the color-rotation distribution of this subsample can be described by the slow sequence with an age of $118_{-16}^{+18}$ Myr. 

We obtained a small number of spectra of this subsample using the Las Cumbres Observatory \citep{2013PASP..125.1031B} Robotic Echelle Spectrographs (NRES) facilities. The stars that were observed exhibit lithium absorption as expected for their youth. The lithium $6708\,\AA$ equivalent width has been noted in Table~\ref{tab:comove} where available. Future works examining the lithium absorption of this subsample may yield a lithium depletion boundary to confirm this age estimate. 

Of these 21 co-moving stars, only one is a canonical member of AB Doradus \citep{2018ApJ...856...23G}. It is likely that the AB Doradus moving group itself is dispersed and ill defined, and our subset of co-moving and potentially co-evolving stars form part of the extended AB Doradus group. 

\begin{deluxetable*}{rrrrrrrrrrrr}
\tablewidth{0pc}
\tabletypesize{\scriptsize}
\tablecaption{
    Candidate co-moving and co-evolving stars with \thisstar{}
    \label{tab:comove}
}
\tablehead{ \\
    \multicolumn{1}{c}{TIC}   &
    \multicolumn{1}{c}{RA (deg)} &
    \multicolumn{1}{c}{DEC (deg)} &
    \multicolumn{1}{c}{U (\kms)} &
    \multicolumn{1}{c}{V (\kms)} &
    \multicolumn{1}{c}{W (\kms)} &
    \multicolumn{1}{c}{B (mag)} &
    \multicolumn{1}{c}{V (mag)} &
    \multicolumn{1}{c}{T (mag)} &
    \multicolumn{1}{c}{K (mag)} &
    \multicolumn{1}{c}{$P_\mathrm{rot}$ (d)} &
    \multicolumn{1}{c}{Li EW $(\AA)$} 
}
\startdata
71314712 & 330.63261 & -47.67748 & -4.0 & -26.0 & -4.0 & 9.3 & 8.8 & 8.2 & 7.4 & 2.4 & \\
91231096 & 310.66707 & -46.67197 & -1.9 & -26.3 & -1.9 & 11.0 & 10.0 & 9.1 & 7.6 & 4.6 & \\
93839949 & 176.15935 & -49.41756 & -7.3 & -24.7 & -7.3 & 9.9 & 8.9 & 8.0 & 6.5 & 1.9 & \\
101403239 & 300.79541 & -52.96796 & -1.7 & -26.0 & -1.7 & 12.8 & 11.9 & 10.9 & 9.4 & 12.7 & \\
108272865 & 285.27521 & -28.71442 & -5.2 & -27.4 & -5.2 & 9.0 & 8.5 & 8.0 & 7.2 & 2.4 & \\
142144969 & 99.73088 & -74.42526 & -5.2 & -26.1 & -5.2 & 10.5 & 9.7 & 9.0 & 7.9 & 7.0 & \\
197597944 & 327.20235 & -39.48626 & -8.5 & -28.1 & -8.5 & 10.3 & 9.7 & 9.0 & 8.0 & 4.1 & \\
206603521 & 21.80129 & -57.29366 & -7.9 & -26.8 & -7.9 & 14.5 & 13.0 & 11.1 & 8.8 & 3.8 & \\
234299476 & 358.66836 & -60.85971 & -8.3 & -27.2 & -8.3 & 11.1 & 10.0 & 9.0 & 7.5 & 9.0 & \\
270200832 & 331.81221 & -74.08641 & -3.0 & -26.4 & -3.0 & 9.5 & 9.0 & 8.5 & 7.8 & 2.0 & \\
270259954 & 359.04562 & -39.05314 & -7.5 & -27.7 & -7.5 & 9.2 & 8.2 & 7.3 & 5.9 & 7.8 & \\
270377865 & 345.08089 & -26.15444 & -3.1 & -26.3 & -3.1 & 8.1 & 7.5 & 6.9 & 5.9 & 3.6 & \\
278271178 & 112.29658 & -82.20387 & -4.4 & -24.9 & -4.4 & 14.3 & 13.1 & 11.1 & 8.9 & 10.9 & \\
280683734 & 342.73660 & -79.16594 & -7.5 & -27.0 & -7.5 & 9.6 & 9.0 & 8.4 & 7.6 & 2.6 & 0.136\\
290081380 & 319.52174 & -29.52160 & -7.0 & -29.4 & -7.0 & 15.1 & 13.9 & 11.5 & 9.3 & 10.3 & \\
341498715 & 318.52194 & -63.70066 & -6.9 & -27.1 & -6.9 & 10.9 & 10.1 & 9.4 & 8.2 & 5.9 &\\
357709300 & 189.06702 & -79.52628 & -7.3 & -26.8 & -7.3 & 12.2 & 11.0 & 10.2 & 8.7 & 7.0 &\\
369897885 & 171.32165 & -84.95449 & -8.4 & -25.4 & -9.6 & 8.1 & 7.6 & 7.2 & 6.5 & 1.5 & 0.030 \\
389660501 & 321.33728 & -43.51207 & -3.5 & -25.8 & -3.5 & 13.4 & 12.1 & 11.0 & 9.1 & 7.1 &\\
405077613 & 112.74683 & -84.32411 & -8.6 & -27.2 & -8.6 & 10.8 & 10.0 & 9.2 & 7.9 & 4.9 & 0.200 \\
409141582 & 312.08079 & -28.02455 & -5.7 & -27.3 & -5.7 & 13.7 & 13.6 & 11.4 & 9.1 & 2.7 &\\
409141582 & 312.08194 & -28.02453 & -9.2 & -25.0 & -9.2 & 13.7 & 13.6 & 11.4 & 9.1 & 2.7 &\\
439417806 & 301.15247 & -35.21447 & -3.7 & -26.2 & -3.7 & 9.4 & 8.9 & 12.2 & 7.7 & 1.4 &\\
464646604 & 287.74114 & -60.27264 & -5.7 & -27.0 & -5.7 & 8.9 & 8.3 & 7.8 & 6.9 & 2.2 & 0.141\\
466277708 & 305.70238 & -65.25823 & -6.9 & -27.0 & -6.9 & 10.3 & 9.6 & 9.0 & 8.1 & 3.8 &\\
\enddata
\end{deluxetable*}


\subsection{X-ray}

The rapid rotation in young stars leads to increased chromospheric activity. As a result, young stars often exhibit higher X-ray and UV emissions than their slowly rotating older counterparts. \thisstar{} is cataloged in the Second ROSAT All-sky Bright Source Catalog \citep{2016AA...588A.103B}. \thisstar{} matches with 2RXS J191057.9-601611, with a count rate of $0.1766\pm0.0425$ counts/second and a hardness ratio of $-0.124\pm0.176$. Using the calibration from \citet{1995ApJ...450..401F}, we find an X-ray luminosity for \thisstar{} of $\log(L_X/L_\mathrm{bol}) = -3.93\pm0.13$. 

The strength of the X-ray activity can help yield a qualitative age estimate for a single star. We adopt Equation A3 from \citet{2008ApJ...687.1264M} to find an approximate age of $50-130$ Myr $(1\sigma)$ for \thisstar{} from its X-ray luminosity, consistent with the expected age of the AB Doradus group.

\subsection{Lithium}
\label{sec:lithium}

Stars undergo lithium depletion during their main-sequence evolution. Lithium is depleted during proton collisions in the cores of stars. Convective mixing between the envelope and the core leads to a gradual depletion of lithium absorption in the observed spectra of Sun-like stars. As such, the lithium abundance of a star can be a tracer for its youth, though direct age estimates from the lithium absorption strength is only qualitative, as is the case for any age indicator when interpreting single stars. Our CHIRON spectra of \thisstar{} reveal significant absorption about the 6708\,\AA{} Li doublet. We model the doublet and the nearby Fe I line simultaneously, measuring a Li equivalent width of $0.1413\pm0.0092\,\AA$ for \thisstar{}. 

\begin{figure}
    \centering
    \includegraphics[width=0.9\linewidth]{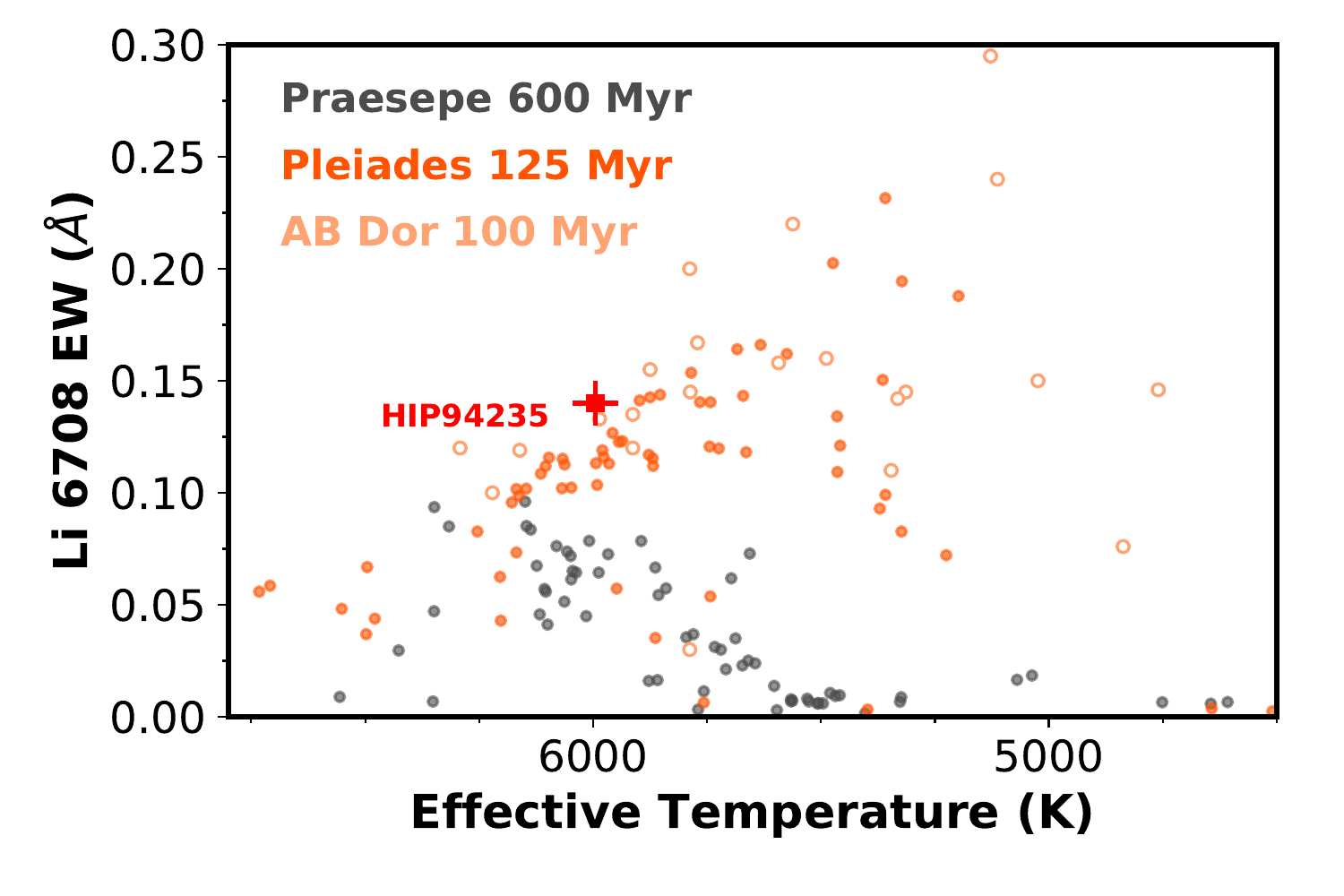}\\
    \includegraphics[width=0.9\linewidth]{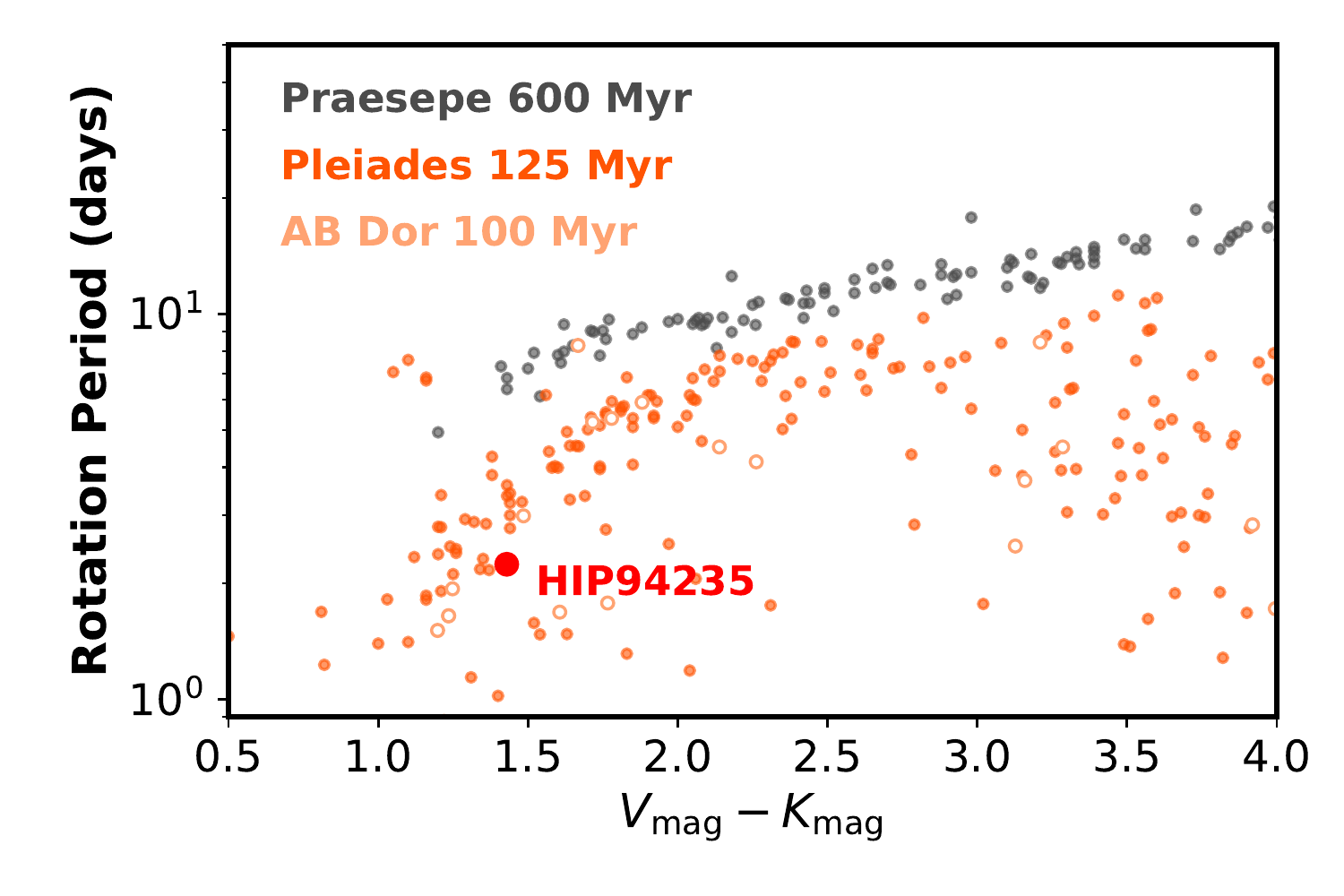}
    \caption{Age indicators of \thisstar{} are consistent with that from AB Doradus and Pleiades members. AB Doradus members are shown in open orange circles, Pleiades members in closed orange circles, and Praesepe members in grey. \textbf{Top} panel shows the distribution of Li 6708\,\AA{} doublet equivalent widths for AB Doradus \citep{2009A&A...508..833D}, Pleiades and Praesepe members \citep{2012ApJ...756L..33Q,2014ApJ...787...27Q,2021AJ....161....2Z}. \textbf{Bottom} panel shows the distribution of rotation periods for the same set of stars.}
    \label{fig:Li}
\end{figure}

Figure \ref{fig:Li} compares the lithium 6708\,\AA{} equivalent width against distributions from known AB Doradus members \citep{2009A&A...508..833D}, Pleiades, and Praesepe clusters \citep{2012ApJ...756L..33Q,2014ApJ...787...27Q,2021AJ....161....2Z}. \thisstar{} is consistent with that of other known members of AB Doradus and Pleiades, and exhibits convincingly stronger lithium absorption than members of the 600\,Myr Praesepe cluster, in agreement with a 120\,Myr age estimate.

\subsection{Rotation}
\label{sec:rotation}

\begin{figure*}
    \centering
    \includegraphics[width=0.45\linewidth]{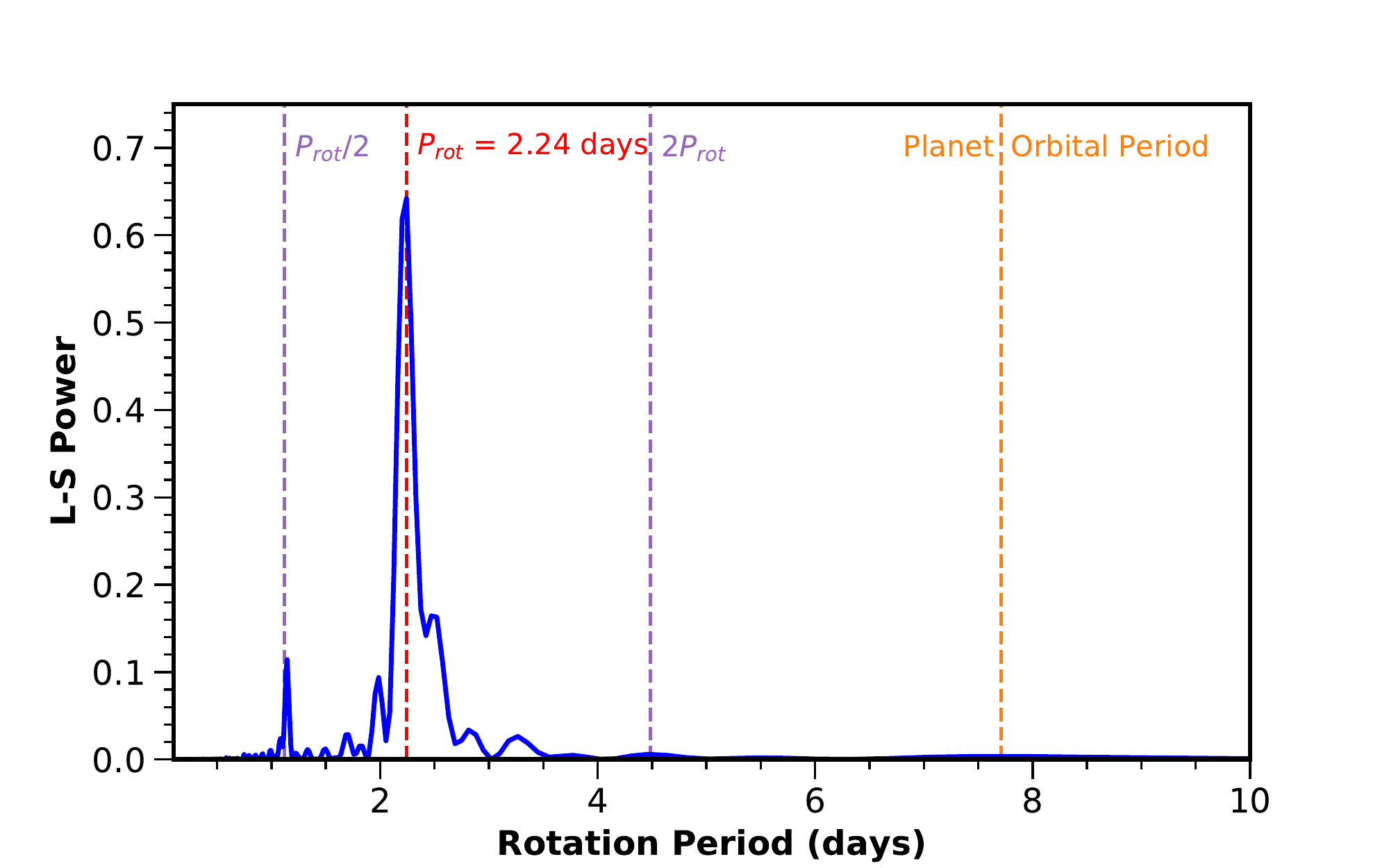}
    \includegraphics[width=0.45\linewidth]{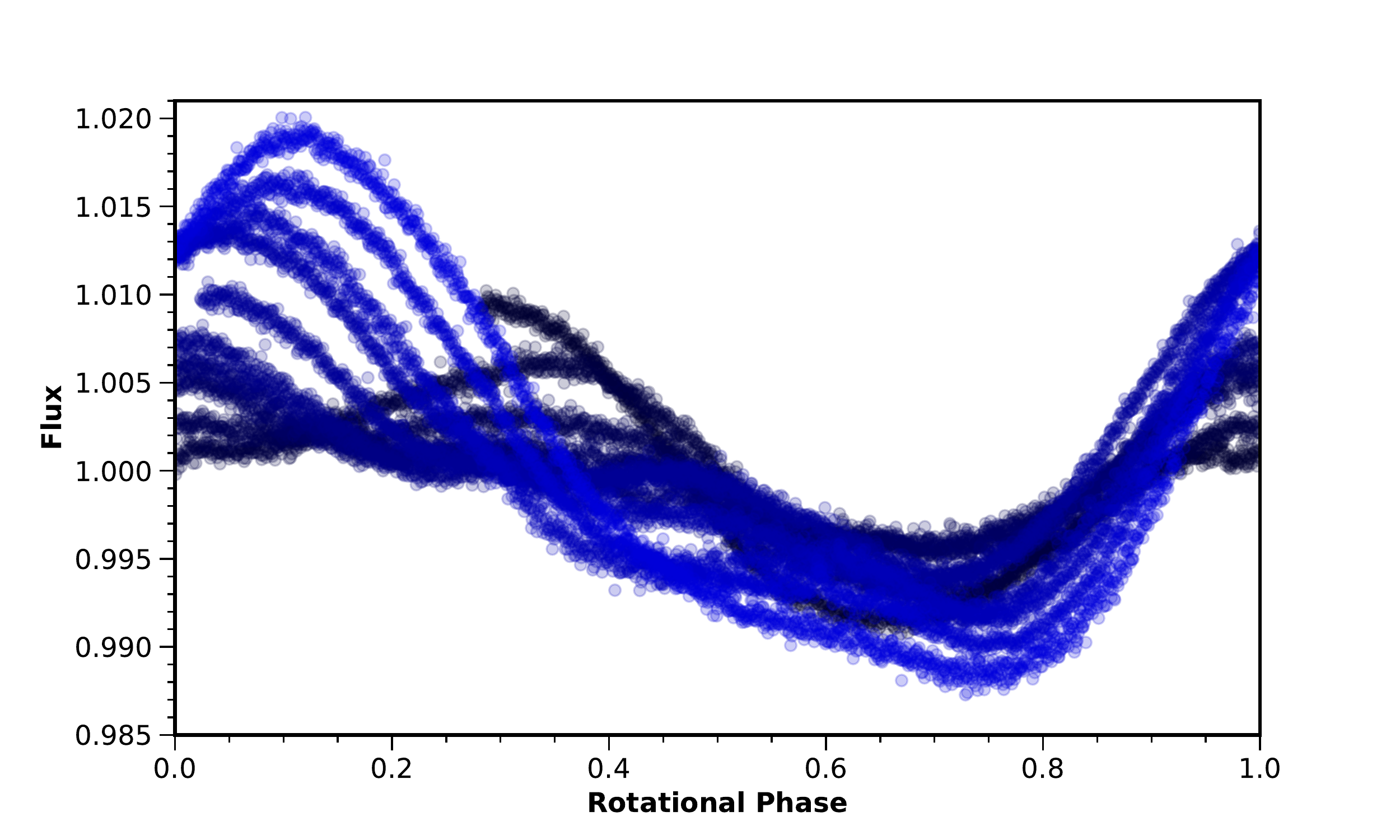}
    \caption{\textbf{Left:} Lomb-Scargle periodogram of the \thisstar{} from its \emph{TESS} light curve. The red dashed line denotes the adopted rotation period, while the dotted purple lines denote the aliases at $P/2$ and $2P$. The orbital period of the planet is marked in orange. \textbf{Right:} Phase-folded TESS light curve, with color gradient applied such that later periods are a lighter blue.}
    \label{fig:LS}
\end{figure*}

The \emph{TESS} observations show that \thisstar{} exhibits significant photometric variability at the 2\% level, consistent with the semi-periodic signature of spot modulation. Figure~\ref{fig:LS} shows the periodicity of \thisstar{} via a Lomb-Scargle periodogram, with a peak rotation period of $2.24\pm0.11$ days. We adopt the width of the rotational peak of the periodogram as the uncertainty on the measured period. The rotational modulation is clear in phase-folded light curve in the right panel of Figure \ref{fig:LS}, and the evolution of starspots over successive rotations can also be seen. Additionally, we modeled the light curve via a stochastically-driven damped harmonic oscillator through a Gaussian process using the \textsc{celerite} package \citep{celerite}. Modeling the posterior via a Markov chain through \textsc{emcee} \citep{2013PASP..125..306F}, we found a posterior distribution for the frequency term $\log \omega_0$ to be $-0.791_{-0.045}^{+0.047}$, corresponding to a period of $2.20 \pm 0.10$ days. 

Figure~\ref{fig:Li} shows the rotation periods of AB Doradus, as well as the Pleiades and Praesepe cluster members. We adopt the membership list for AB Doradus from \citet{2018ApJ...856...23G}, and derived rotation periods for stars with available \emph{TESS} light curves from the MIT FFI QLP library \citep{2020RNAAS...4..206H} for stars with available light curves. Rotation periods are measured via a Lomb-Scargle period search, and down-selected by hand to remove stars that do not show unambiguous rotational signatures. By-hand corrections of period aliases were also applied. Of the 66 members listed in \citet{2018ApJ...856...23G}, 55 were bright enough and had accessible \emph{TESS} QLP light curves, and 37 yielded convincing detections of a rotation signal. Rotation periods for Pleiades members were adopted from \citet{2016AJ....152..113R}, and for Praesepe from \citet{2017ApJ...839...92R}. The rotation of \thisstar{} agrees with the Pleiades and AB Doradus distribution, in agreement with its kinematic age estimates. 


\begin{deluxetable*}{lrrrr}
\tablewidth{0pc}
\tabletypesize{\scriptsize}
\tablecaption{
    Properties of \thisstar{}
    \label{tab:stellar}
}
\tablehead{ \\
    \multicolumn{1}{c}{~~~~~~~~Parameter~~~~~~~~}   &
    \multicolumn{1}{c}{Value} &
    \multicolumn{1}{c}{Source} &
}
\startdata
\sidehead{\textbf{Astrometry}}
~~~Right Ascension \dotfill & 19:10:57.87 & \citet{2021AA...649A...1G}\\
~~~Declination \dotfill & $-$60:16:21.49 & \citet{2021AA...649A...1G}\\
~~~Parallax (mas) \dotfill & $17.061\pm0.037$ & \citet{2021AA...649A...1G}\\
\sidehead{\textbf{Proper Motion}}
~~~Gaia (2016.1) RA Proper Motion (mas yr$^{-1}$) \dotfill & $11.632\pm0.025$ & \citet{2021AA...649A...1G}\\
~~~Gaia (2016.1) Dec Proper Motion (mas yr$^{-1}$) \dotfill & $-100.836\pm0.025$ & \citet{2021AA...649A...1G}\\
~~~Hipparcos (1991.1) RA Proper Motion (mas yr$^{-1}$) \dotfill & $12.755\pm0.890$ & \citet{1997ESASP1200.....E}\\
~~~Hipparcos (1991.4) Dec Proper Motion (mas yr$^{-1}$) \dotfill & $-99.617\pm0.796$ & \citet{1997ESASP1200.....E}\\
~~~Hipparcos-Gaia Average RA Proper Motion (mas yr$^{-1}$) \dotfill & $11.509\pm0.025$ & \citet{2021ApJS..254...42B}\\
~~~Hipparcos-Gaia Dec Average Proper Motion (mas yr$^{-1}$) \dotfill & $-100.628\pm0.019$ & \citet{2021ApJS..254...42B}\\
\sidehead{\textbf{Photometry}}
~~~TESS (mag) \dotfill & $7.758\pm0.006$ & \citet{2018AJ....156..102S}\\
~~~B (mag) \dotfill & $8.943\pm0.027$ & \citet{2016yCat.2336....0H}\\
~~~V (mag) \dotfill & $8.31\pm0.03$ & \citet{2016yCat.2336....0H}\\
~~~J (mag) \dotfill & $7.201\pm0.023$ & \citet{2006AJ....131.1163S}\\
~~~H (mag) \dotfill & $6.966\pm0.023$ & \citet{2006AJ....131.1163S}\\
~~~K (mag) \dotfill & $6.881\pm0.027$ & \citet{2006AJ....131.1163S}\\
~~~\emph{Gaia} (mag) \dotfill & $8.173207\pm0.00031$ & \citet{2021AA...649A...1G}\\
~~~$\mathrm{\emph{Gaia}}_\mathrm{BP}$ (mag) \dotfill & $8.4634\pm0.0013$ & \citet{2021AA...649A...1G}\\
~~~$\mathrm{\emph{Gaia}}_\mathrm{RP}$ (mag) \dotfill & $7.70637\pm0.00087$ & \citet{2021AA...649A...1G}\\
~~~WISE W1 (mag) \dotfill & $6.841\pm0.066$ & \citet{2012yCat.2311....0C}\\
~~~WISE W2 (mag) \dotfill & $6.823\pm0.02$ & \citet{2012yCat.2311....0C}\\
~~~WISE W3 (mag) \dotfill & $6.836\pm0.016$ & \citet{2012yCat.2311....0C}\\
~~~WISE W4 (mag) \dotfill & $6.728\pm0.067$ & \citet{2012yCat.2311....0C}\\
\sidehead{\textbf{Kinematics and Position}}
~~~$U$ ($\kms$) \dotfill & $-5.61\pm0.34$ & Derived\\
~~~$V$ ($\kms$) \dotfill & $-27.03\pm0.17$ & Derived\\
~~~$W$ ($\kms$) \dotfill & $-11.82\pm0.18$ & Derived\\
~~~Distance (pc)       \dotfill    & $58.54_{-0.07}^{+0.08}$ & Derived\\
\sidehead{\textbf{Physical Properties}}
~~~$M_\star$ ($M_\odot$)      \dotfill    &  $1.094_{-0.007}^{+0.024}$ & Fitted (Uniform Prior)\\
~~~$R_\star$ ($R_\odot$)      \dotfill    &  $1.08_{-0.10}^{+0.11}$ & Fitted (Uniform Prior)\\
~~~Effective Temperature $T_\mathrm{eff}$ (K) \dotfill    &  $5991\pm50$ & This paper\\
~~~Surface Gravity $\log g_\star$ (cgs) \dotfill    &  $4.460\pm0.05$ & This paper\\
~~~[m/H] \dotfill & 0.0 & \citep{2013ApJ...766....6B} \\
~~~$v\sin I_\star$ $(\kms)$ \dotfill & $24.4\pm1.0$ & This paper \\
~~~$I_\star\,(^\circ)$ \dotfill & $>70$ $(3\sigma)$ & Calculated as per \citet{2020AJ....159...81M} \\
~~~Age (Myr) \dotfill & 50-150 & \citet{2004ApJ...613L..65Z}\\
    & & \citet{2005ApJ...628L..69L}\\
    & & \citet{2015MNRAS.454..593B}\\
~~~Limb darkening coefficients (TESS) \dotfill & (0.17, 0.41) & \citet{2017AA...600A..30C}\\
~~~Limb darkening coefficients (CHEOPS) \dotfill & (0.27, 0.29) & \citet{2017AA...600A..30C}\\
\sidehead{\textbf{Activity indicators}}
~~~$P_{rot}$ (days) \dotfill & $2.24\pm0.11$ & This paper\\
~~~Lithium $6708\,\AA$ Equivalent Width $(\AA)$ \dotfill & $0.1413\pm0.0092$ & This paper\\
~~~X-Ray luminosity $\log(L_X/L_\mathrm{bol})$ & $-3.93\pm0.13$ & \citep{2016AA...588A.103B}\\
\enddata
\end{deluxetable*}

\section{Global model}
\label{sec:analysis}

To best estimate the stellar and planetary properties of the \thisstar{} system, we perform a global modeling incorporating the \emph{TESS} and \emph{CHEOPS} transits, radial velocities, and photometric and spectroscopic properties of \thisstar{}. 

The transit models were computed as per \citet{2002ApJ...580L.171M}, implemented via \textsc{batman} \citep{2015PASP..127.1161K}. The free parameters for this model are the stellar mass and radius, the time of transit center $T_0,$ the orbital period $P$, the planetary radius ratio $R_p/R_\star$, and the eccentricity parameters $\sqrt{e}\cos{\omega}$ and $\sqrt{e}\sin{\omega}$, where $e$ is the eccentricity and $\omega$ is the longitude of periastron. 

We model the \emph{CHEOPS} transit and its associated stellar variability and instrumental characteristics simultaneous to the global modeling. The transit model is computed as per \citet{2002ApJ...580L.171M}. The hours-long rotational modulation signal is modeled via a 4th degree polynomial with respect to time. The correlation between the light curve and the spacecraft motion is modeled via a 5th degree polynomial against the roll angle. Figure \ref{fig:cheopsmodel} shows the \emph{CHEOPS} light curve before and after the removal of the best fit stellar variability and instrumental model. 

The out-of-transit Keplerian radial velocity was modeled with additional parameters describing the systemic velocity $\gamma$, planetary mass $\mpl$, and a jitter term for each instrument. 

The stellar mass and radius were modeled using the MIST isochrones \citep{2016ApJS..222....8D}, and constrained by their photometric magnitudes and parallax priors from \emph{Gaia} \emph{G}, {Bp}, {Rp} \citep{2018AA...616A...1G}, \emph{Hipparcos} TYCHO $B$ and $V$ bands \citep{1997AA...323L..49P}, 2MASS $J$, $H$, and $Ks$ bands \citep{2006AJ....131.1163S}. Additionally, the age was restricted to be $120\pm50$ Myr as per the age of the AB Dor moving group, and the limb darkening coefficients were fixed to theoretically interpolated values \citep{Claret:2011,2017AA...600A..30C}. All other parameters were assigned uniform priors with physically motivated boundaries.

\begin{figure}
    \centering
    \includegraphics[width=\linewidth]{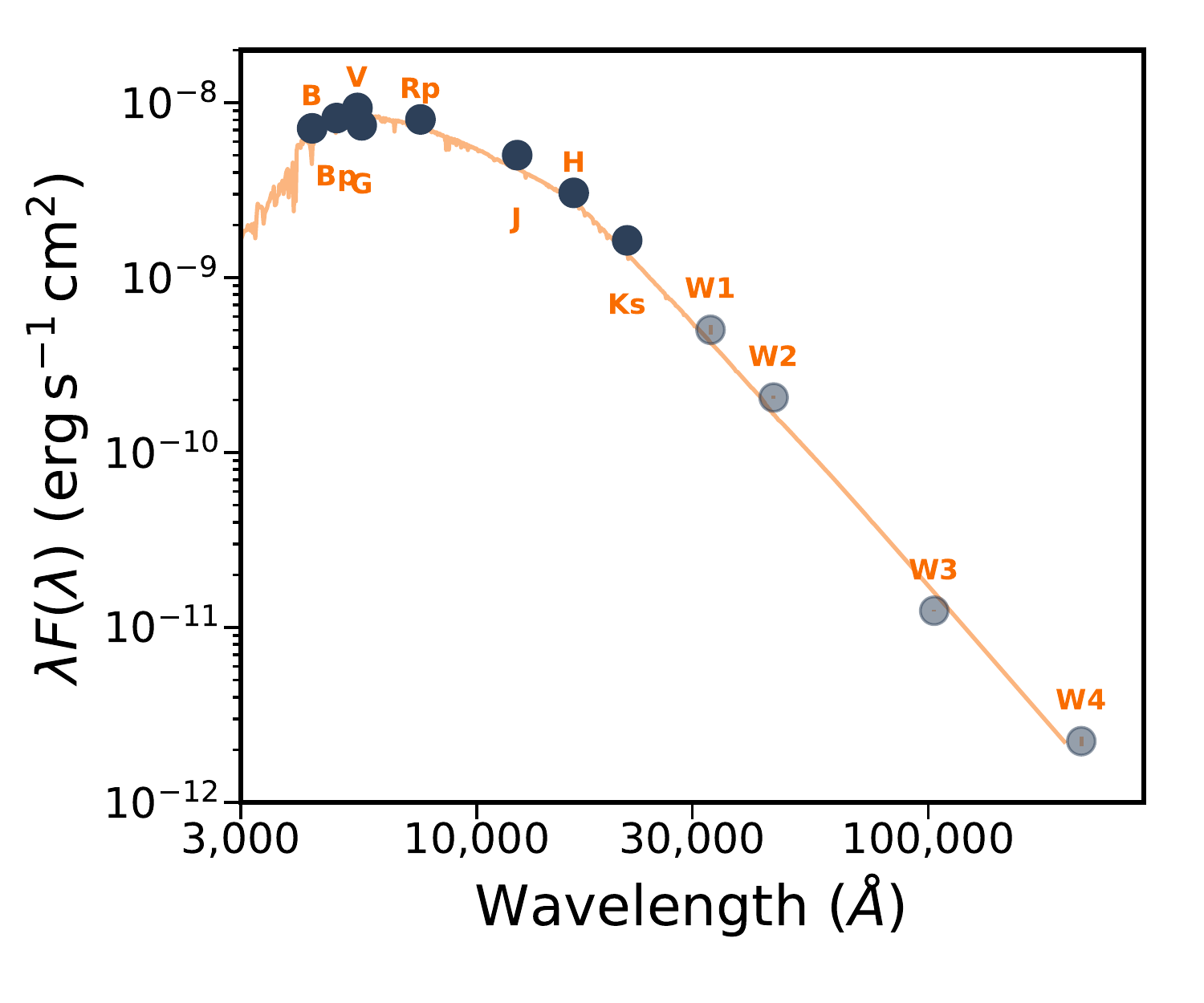}
    \caption{Spectral energy distribution of \thisstar{}. Magnitudes from Tycho $B$, $V$, Gaia $Bp$, $G$, $Rp$, 2MASS $J$, $H$, $Ks$, and WISE $W1$, $W2$, $W3$,$W4$ are plotted. We note that WISE $W1$, $W2$, $W3$,$W4$ were not used in the global modeling, and are marked by a lighter color compared to the bluer bands. The template spectrum from ATLAS9 model atmospheres \citep{Castelli:2004} is plotted for reference.}
    \label{fig:sed}
\end{figure}

The modeling was performed simultaneously for all parameters using Monte Carlo Markov Chain (MCMC) analysis, making use of the \texttt{emcee} package \citep{2013PASP..125..306F}. Results are listed in Tables \ref{tab:stellar} and \ref{tab:planet}, the best fit light curve model is shown in Figure \ref{fig:lightcurve}, and the best fit radial velocity model is shown in Figure \ref{fig:rvs}. The spectral energy distribution of \thisstar{} is shown in Figure~\ref{fig:sed}, along with the template ATLAS9 model spectrum \citep{Castelli:2004} computed at the best fit stellar parameters of \thisstar{}. We note that the mid-infrared WISE magnitudes show no excess that might be indicative of a remnant debris disk around the young star. 

\begin{deluxetable*}{lrrrr}
\tablewidth{0pc}
\tabletypesize{\scriptsize}
\tablecaption{
    Derived parameters for \thisstarb{}
    \label{tab:planet}
}
\tablehead{ \\
    \multicolumn{1}{c}{~~~~~~~~Parameter~~~~~~~~}   &
    \multicolumn{1}{c}{Joint model} &
    \multicolumn{1}{c}{Priors} &
}
\startdata
\sidehead{\textbf{Fitted Parameters}}
~~~$T_0$ (BJD)              \dotfill & $2459037.8704_{-0.0022}^{+0.0011}$ & Uniform\\
~~~$P$ (days)             \dotfill    &  $ 7.713057_{-0.000021}^{+0.000021}$ & Uniform \\
~~~$\rpl/\rstar$ \dotfill & $0.0253_{-0.00059}^{+0.00075}$ & Uniform\\
~~~$i$ (deg) \dotfill & $87.14_{-0.17}^{+0.16}$ & Uniform\\
~~~$\sqrt{e}\cos{\omega}$      \dotfill    &  $0.07_{-0.54}^{+0.50}$ & Uniform\\
~~~$\sqrt{e}\sin{\omega}$      \dotfill    &  $0.26_{-0.27}^{+0.20}$ & Uniform\\
~~~$\mpl$ ($\mearth$)      \dotfill    &  $<379$ $(3\sigma)$& Uniform\\
~~~$\gamma_{\mathrm{MINERVA-Australis 3}}$ \dotfill & $9013_{-70}^{+58}$ & Uniform\\
~~~$\gamma_{\mathrm{MINERVA-Australis 5}}$ \dotfill & $8960_{-126}^{+123}$ & Uniform\\
~~~$\gamma_{\mathrm{MINERVA-Australis 6}}$ \dotfill & $9050_{-81}^{+88}$ & Uniform\\
~~~$\gamma_{\mathrm{CHIRON}}$ \dotfill & $8264_{-30}^{+31}$ & Uniform\\
~~~RV Jitter MINERVA-Australis 3 \dotfill & $80_{-56}^{+73}$ & Uniform\\
~~~RV Jitter MINERVA-Australis 5 \dotfill & $154_{-98}^{+210}$ & Uniform\\
~~~RV Jitter MINERVA-Australis 6 \dotfill & $140_{-79}^{+113}$ & Uniform\\
~~~RV Jitter CHIRON \dotfill & $100_{-28}^{+35}$ & Uniform\\
\sidehead{\textbf{Inferred parameters}}
~~~$e$      \dotfill    &  $0.32_{-0.20}^{+0.20}$ & Derived\\
~~~$\omega$ (deg)    \dotfill    &  $17_{-92}^{+67}$ & Derived\\
~~~$\rpl$ ($R_\oplus$)       \dotfill    & $3.00_{-0.28}^{+0.32}$ & Derived\\
~~~$a/R_\star$ \dotfill & $15.7_{-1.5}^{+1.6}$ & Derived\\
~~~$a$ ($AU$) \dotfill & $0.07870_{-0.00017}^{+0.00056}$ & Derived\\
~~~Transit duration (days) \dotfill & $0.103_{-0.018}^{+0.009}$ & Derived\\
~~~$K_\mathrm{rv}$ $(\ms)$ \dotfill & $<150$ $(3\sigma)$ & Derived\\
~~~$T_\mathrm{eq}$ (K) \dotfill & $1060\pm50$ & Derived\\
\enddata
\end{deluxetable*}

\section{\thisstar{} B: A 60 AU M-dwarf companion}

\begin{deluxetable*}{llrrrl}
\tablewidth{0pc}
\tabletypesize{\scriptsize}
\tablecaption{
    Diffraction limited measurements of \thisstar{} B
    \label{tab:AO}
}
\tablehead{ \\
    \multicolumn{1}{c}{Instrument}   &
    \multicolumn{1}{c}{Epoch}   &
    \multicolumn{1}{c}{Separation} &
    \multicolumn{1}{c}{Position Angle} &
    \multicolumn{1}{c}{$\Delta m$} &
    \multicolumn{1}{c}{Reference} \\
    \multicolumn{1}{c}{}   &
    \multicolumn{1}{c}{}   &
    \multicolumn{1}{c}{(mas)} &
    \multicolumn{1}{c}{($^\circ$)} &
    \multicolumn{1}{c}{(mag)} &
    \multicolumn{1}{c}{} 
}
\startdata
VLT-NaCo & 2010-07-30 & $506 \pm 7$ & $150.6 \pm 0.8$  & $3.8\pm0.3$ (H band) & \citet{2015AA...573A.127C}, \citet{2015AA...573A.126D}\\
Gemini-Zorro & 2021-07-23 & $596\pm5$ & $162.87\pm0.48$  & 5.84 (832 nm) & This Work\\
Gemini-Zorro & 2021-10-22 & $600\pm8$ & $161.73\pm0.75$  & 5.31 (832 nm) & This Work\\
\enddata
\end{deluxetable*}

Diffraction limited observations revealed a bound M-dwarf companion to \thisstar{}. The companion was identified in speckle imaging observations of \thisstar{} during the candidate vetting process, and also identified from archival adaptive optics observations searching for wide Jovian companions to young stars 11 years prior. We describe the observations below, and perform compute astrometric orbital solutions for \thisstar{}B. 

\begin{figure}
    \centering
    \includegraphics[width=\linewidth]{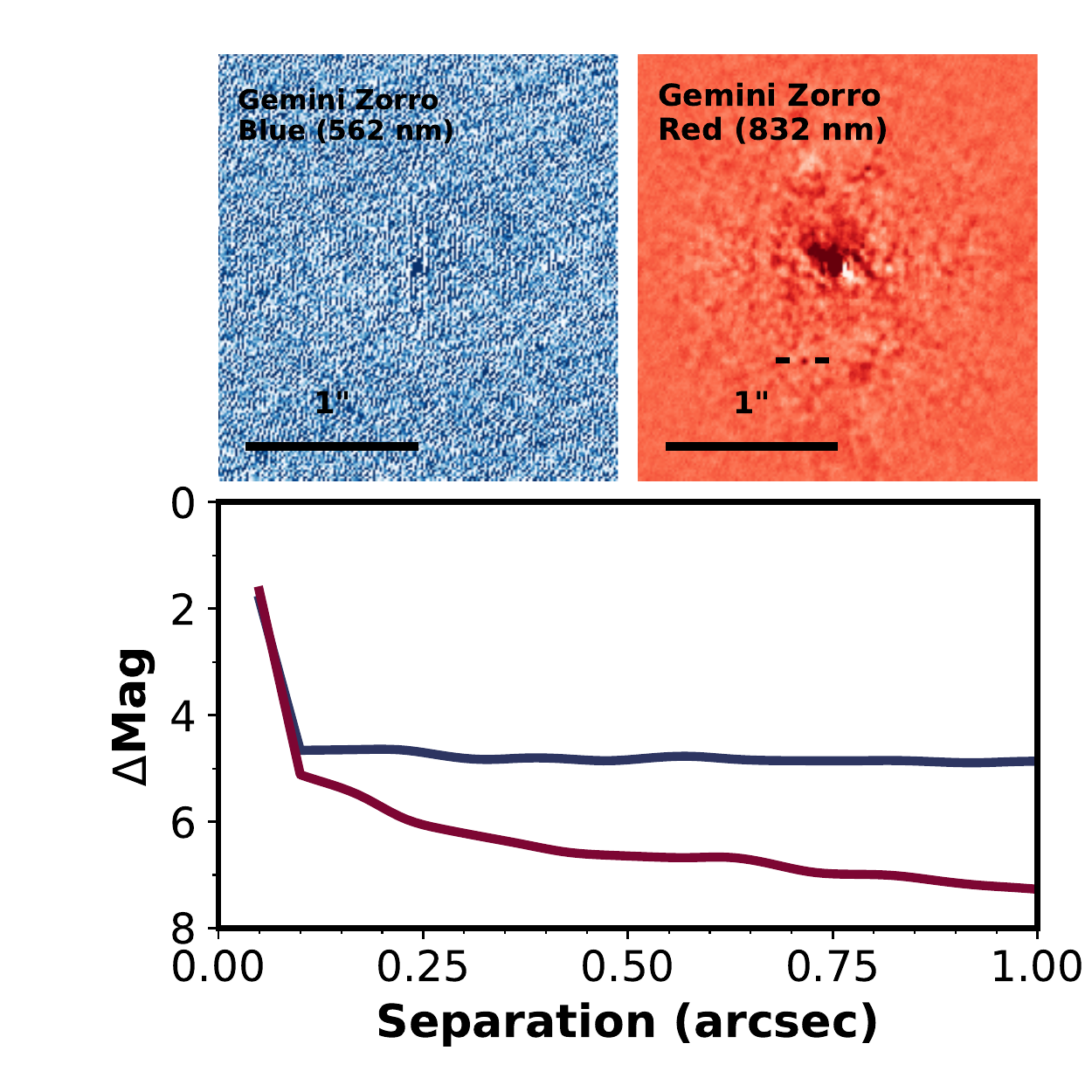}
    \caption{Gemini South Zorro speckle observations of \thisstar{} on 2021-10-22. The speckle auto cross correlation functions are shown in the top row, with the image from the Blue camera at 562nm on the \textbf{left}, and the Red camera at 832nm on the \textbf{right}. A companion is detected in the red arm with a contrast of $\Delta m = 5.3$ at a separation of $0.6\arcsec$. No companions were detected on the blue arm with a limit of $\Delta m > 4.7$. The companion is marked in the red arm image.}
    \label{fig:speckle}
\end{figure}

We obtained high-contrast imaging at 562nm and 832\,nm with the Zorro speckle imager on the 8\,m Gemini South Observatory on 2021-07-23 and 2021-10-22, which revealed a faint stellar companion to \thisstar{}. Zorro is a dual-channel speckle imager with a pixel scale of $0.01\arcsec \mathrm{pixel}^{-1}$ and an approximate full width at half maximum of $0.02\arcsec$. Data reduction and analysis were performed as per \citet{2011AJ....142...19H} and \citet{2016ApJ...829L...2H}. Figure~\ref{fig:speckle} shows the Zorro image at 562nm and 832\,nm of \thisstar{} on 2021-10-22. The 832\,nm observation achieved a contrast ratio of $\Delta m=6.68$ at $0.5\arcsec$ separation. A companion of $\Delta m=5.3$ was identified at a separation of $0.6\arcsec$ in the red band. No companions with contrasts brighter than $\Delta m = 4.7$ were detected in the blue arm. Observations from 2021-07-23 also identified the same companion in the 832\,nm observations, though the blue arm was not functional at the time. 

Adaptive optics imaging of \thisstar{} was also carried out as part of a large program to characterize the occurrence rates of giant planets at large separations by \citet{2015AA...573A.127C}. These observations, using the NaCo high contrast Adaptive Optics (AO) imager on VLT-UT4, were obtained on 2010-07-30. They revealed the stellar companion at a separation of $0.5\arcsec$ with a contrast of $\Delta H = 3.8\pm0.3$. The positional information from the diffraction limited imaging observations are listed in Table~\ref{tab:AO}.

Based on the \emph{Gaia} and \emph{Hipparcos} proper motions measurements, the accumulated motion of \thisstar{} over the 11 year interval is $1.1\arcsec$. In contrast, the relative motion between the imaged companion and the target star is $0.1\arcsec$, which strongly suggests the pair is a a bound stellar binary with a projected separation of 31 AU. We henceforth refer to this companion as \thisstar{}~B.

To determine the approximate properties of \thisstar{}~B, we adopt the 100\,Myr MIST isochrones \citep{2016ApJ...823..102C} to model its 832 nm and $H$-band magnitudes. We approximate the 832nm band with the $I$ band magnitudes from MIST, and adopt magnitudes for the companion of $I=10.76\pm0.3$ and $H=13.05\pm0.4$. We adopt a fixed metallicity of $[\mathrm{M/H}]=0$ as per that for the AB Doradus moving group \citep{2009A&A...508..833D}, and adopt a strong Gaussian prior for the distance to \thisstar{}~B as per its \emph{Gaia} parallax. We find the companion is an M-dwarf of mass $0.26\pm0.04\,M_\odot$ and radius $0.31\pm0.03\,R_\odot$. The companion is incapable of hosting the transiting companion responsible for the planetary transit signal detected in \emph{TESS}. To check if this companion is capable of being the source of the transit signal, we deblend and refit for the transit about the M-dwarf companion. We find a best fit radius ratio of $R_2/R_1 = 0.46\pm0.01$ for such a system, with a V-shaped transit that is incompatible with the observed light curve. A model comparison yields a Bayesian information criterion difference of $\Delta \mathrm{BIC} = 65.9$ between the best fit model of a transit about the companion M-dwarf and that of our nominal planetary transit scenario. We therefore rule out the companion being the source of the observed transit signal.

Although the available imaging detections cover only a small part of the orbital arc of \thisstar{}~B, it is possible in principle that the binary orbit could be constrained. Several authors \citep[e.g.][]{2018A&A...615A.149C, 2018NatAs...2..883S, 2018ApJS..239...31B, 2019A&A...623A..72K} have demonstrated that it is possible to combine proper motion measurements from \emph{Gaia} and \emph{Hipparcos} to produce long-timescale astrometric data that can be used to detect the reflex orbital motion of orbiting companions. \citet{2019AJ....158..140B} extended this technique further by jointly fitting Hipparcos-Gaia astrometry with radial velocities and relative astrometry from direct imaging, and found that the orbits of massive companions can be precisely constrained even when the orbital periods are much longer than the observational duration. This method has been profitably applied to many further systems; \citet{2021AJ....161..106B}, for example, were able to extract precise orbital parameters and masses for two white dwarfs with orbital periods in excess of $\gtrsim200$~yrs despite possessing no more than $\sim30$~yrs of observational data in both cases.

Inspection the Hipparcos-Gaia Catalog of Accelerations \citep{2021ApJS..254...42B} shows that \thisstar{} displays a statistically significant astrometric acceleration of 0.24~mas~yr$^{-1}$ ($\chi^2=58.5$) between the \emph{Gaia} and \emph{Hipparcos-Gaia} proper motions, equivalent to a drift in the stellar tangential velocity of $\approx70~\ms$, which can plausibly be attributed to \thisstar{}~B. This motivates us to attempt an orbital fit for \thisstar{}~B based on the available data.

\begin{deluxetable*}{lrrr}
\tablewidth{0pc}
\tabletypesize{\scriptsize}
\tablecaption{
    Orbital parameters of \thisstar{} B
    \label{tab:binary}
}
\tablehead{ \\
    \multicolumn{1}{c}{Parameter}   &
    \multicolumn{1}{c}{Median $\pm$ 1$\sigma$} &
    \multicolumn{1}{c}{Mode}   &
    \multicolumn{1}{c}{Priors} 
}
\startdata
\sidehead{\textbf{Informed Priors}}
~~~Parallax (mas) \dotfill & $17.061 \pm 0.037$ & - & Gaussian \\
~~~$M_\mathrm{A}$ ($M_\odot$) \dotfill & $1.094 \pm 0.05$ & - & Gaussian \\
~~~$M_\mathrm{B}$ ($M_\odot$) \dotfill & $0.26 \pm 0.04$ & - & Gaussian \\
\sidehead{\textbf{Fitted Parameters}}
~~~$a$ (AU) \dotfill & $56^{+9}_{-7}$ & $54$ & Log-uniform \\
~~~$\sqrt{e}\sin{\omega}$ \dotfill & $0.35^{+0.14}_{-0.19}$ & $0.40$ & Uniform \\
~~~$\sqrt{e}\cos{\omega}$ \dotfill & $0.20^{+0.48}_{-0.34}$ & $0.48$ & Uniform \\
~~~Mean anomaly at BJD~$=2457000$ (deg) \dotfill & $170^{+110}_{-80}$ & $130$ & Uniform \\
~~~$i$ (deg) \dotfill & $67.8^{+2.7}_{-2.9}$ & $67.4$ & $\sin{i}$ \\
~~~$\Omega$ (deg) \dotfill & $20^{+11}_{-7}$ & $18$ & Uniform \\
~~~Barycentric RA proper motion (mas yr$^{-1}$) \dotfill & $10.50 \pm 0.24$ & $10.46$ & Uniform \\
~~~Barycentric declination proper motion (mas yr$^{-1}$) \dotfill & $-102.99 \pm 0.35$ & $102.91$ & Uniform \\
\sidehead{\textbf{Derived Parameters}}
~~~$P$ (years) \dotfill & $365^{+92}_{-69}$ & $362$ & Derived \\
~~~$e$ \dotfill & $0.25^{+0.22}_{-0.14}$ & $0.19$ & Derived \\
~~~$\omega$ (deg) \dotfill & $300^{+30}_{-80}$ & $320$ & Derived \\
~~~Periastron distance (AU) \dotfill & $43^{+11}_{-15}$ & $47$ & Derived \\
~~~Time of periastron (years CE) \dotfill & $2184^{+107}_{-74}$ & $2164$ & Derived \\
\enddata
\end{deluxetable*}

\begin{figure*}
    \centering
    \includegraphics[width=0.45\linewidth]{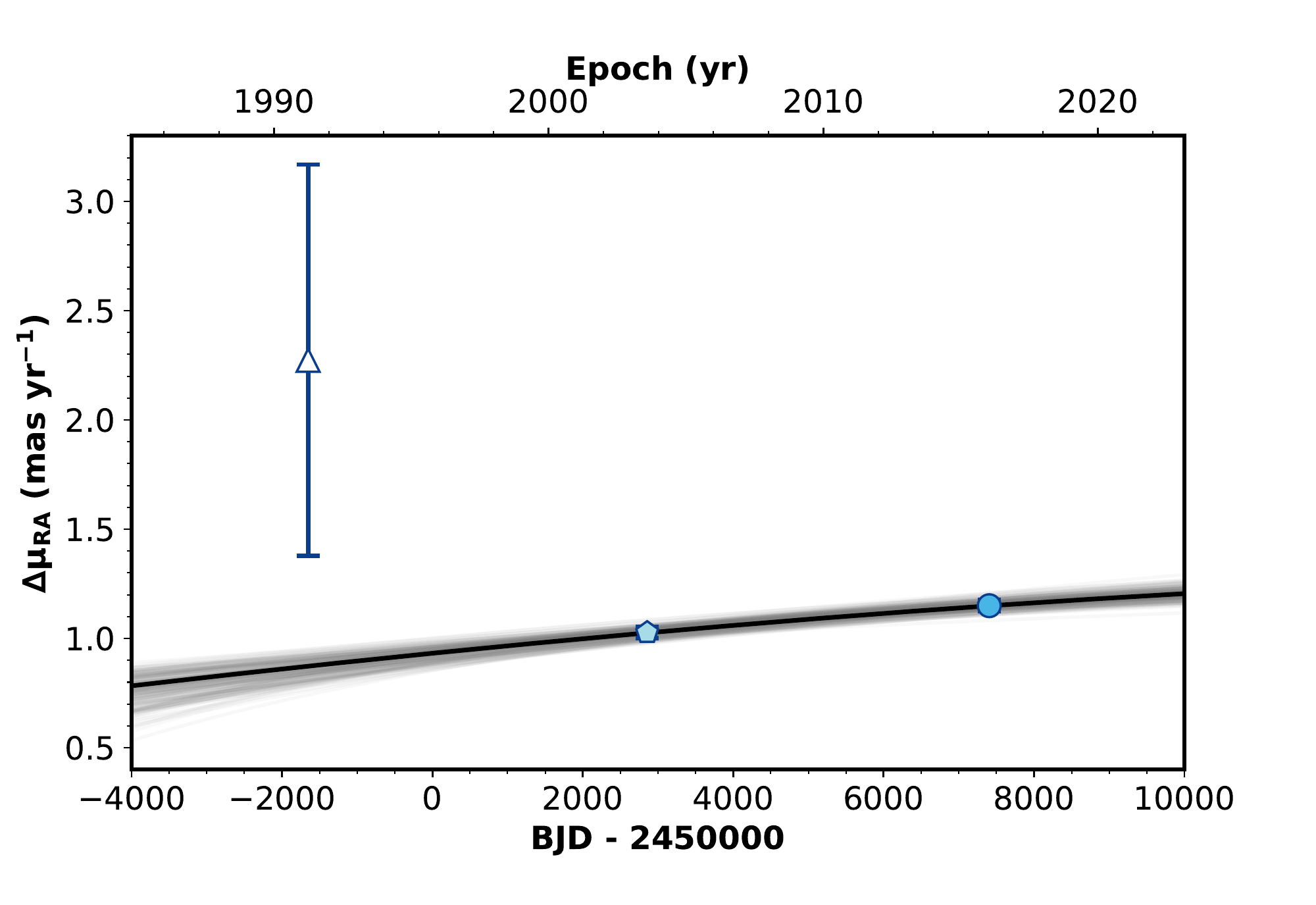}
    \includegraphics[width=0.45\linewidth]{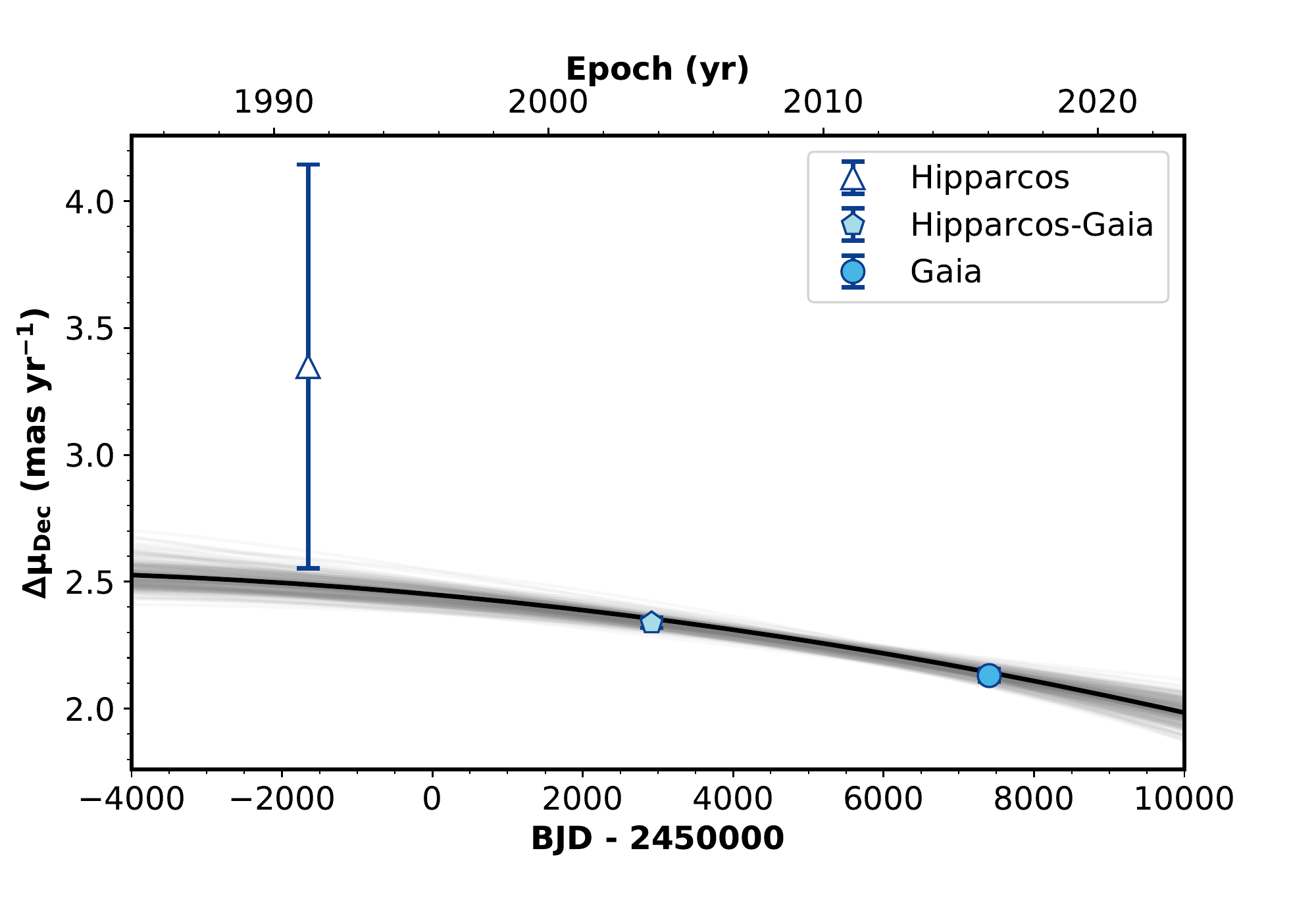}
    \includegraphics[width=0.45\linewidth]{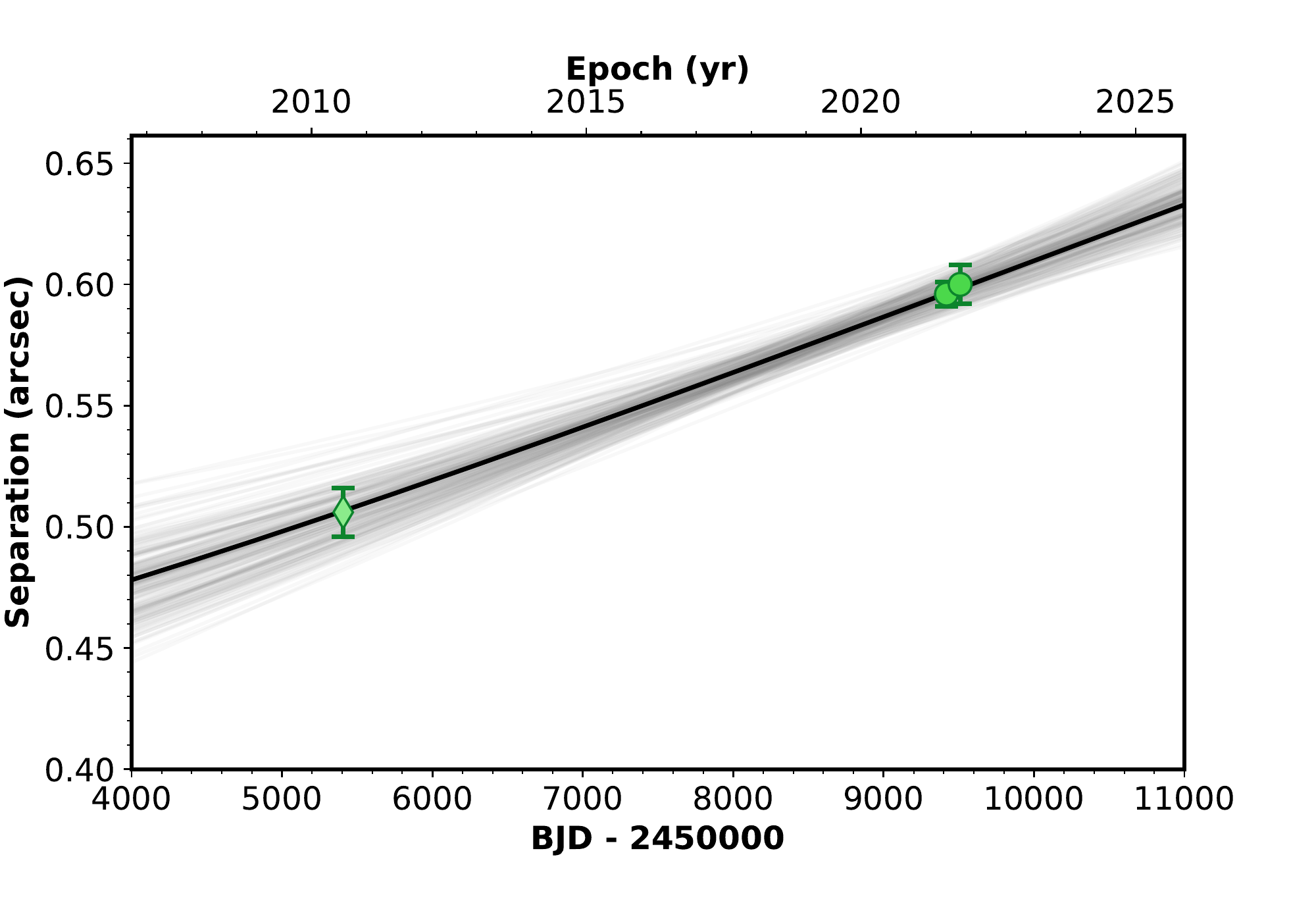}
    \includegraphics[width=0.45\linewidth]{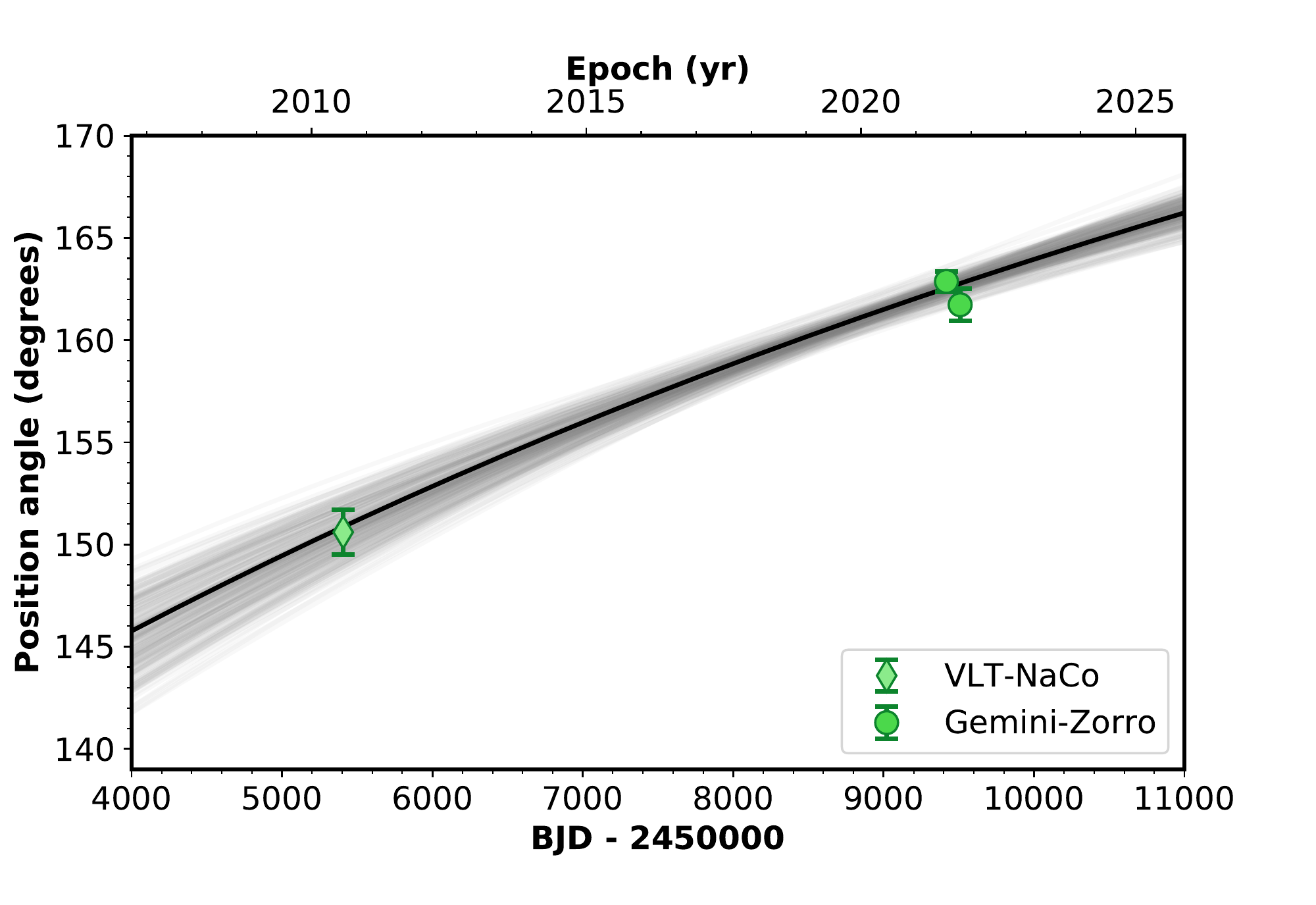}
    \caption{Keplerian orbital model to the proper motion \textbf{of \thisstar{}} (\textbf{top left} in right ascension, \textbf{top right} in declination) and relative astrometry \textbf{of \thisstar{}~B} (\textbf{bottom left} in separation, \textbf{bottom right} in position angle). The best-fit model is shown in black, while the orbits in gray are drawn randomly from the posteriors. In the fit to the proper motions, all values are normalised to the proper motion of the system barycenter.}
    \label{fig:binary_fit}
\end{figure*}

\begin{figure}
    \centering
    \includegraphics[width=\linewidth]{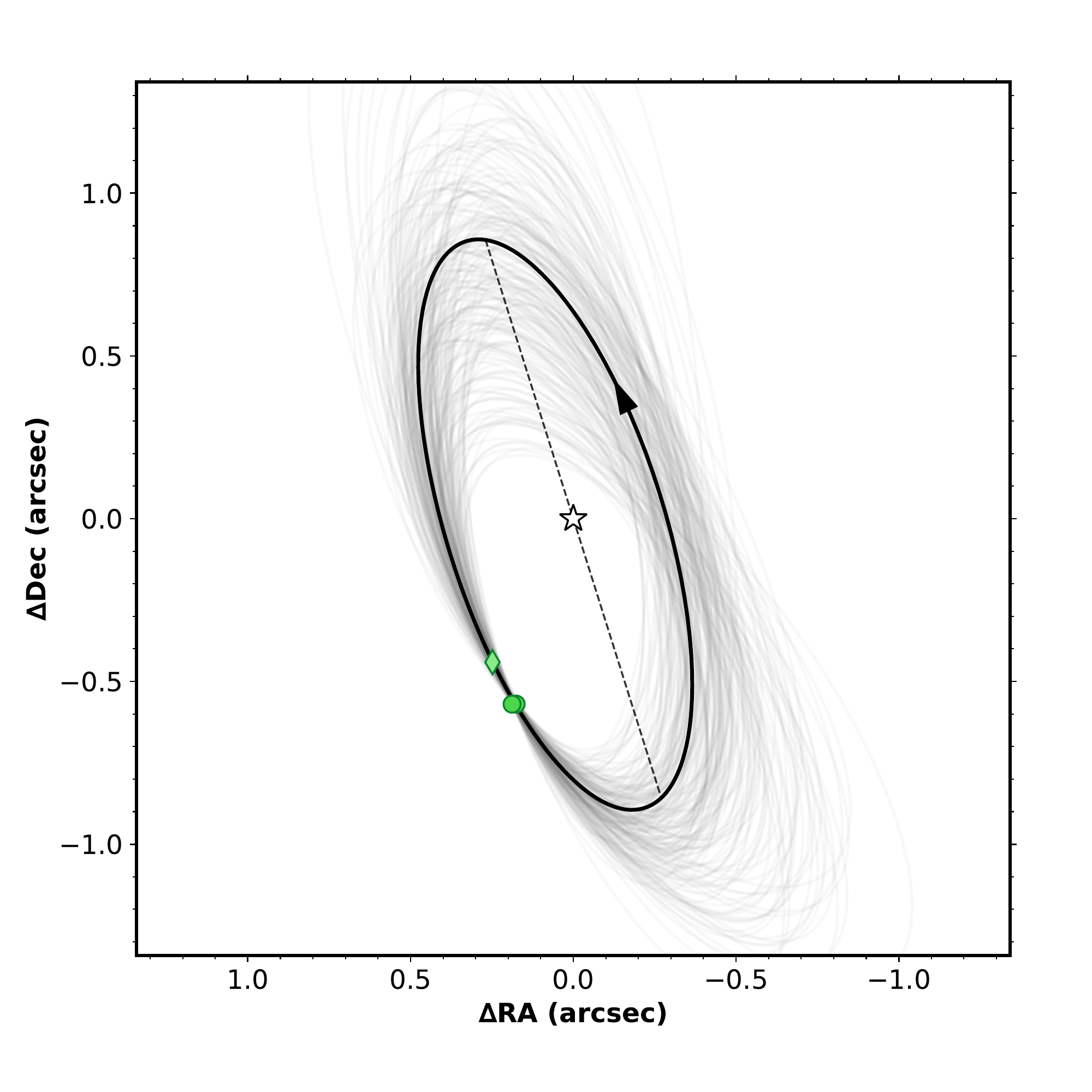}
    \caption{The projected sky orbit of \thisstar{}~B. The data format is as in Figure \ref{fig:binary_fit}. The dotted line marks the intersection of the planes of the orbit and of the sky and the black arrowhead indicates the direction of motion. Despite the short arc of observations our joint fit allows us to robustly constrain several orbital parameters, including strong constraints on the orbital inclination ($i~=~67.8^{+2.7}_{-2.9}$~deg) and semi-major axis ($a~=~56^{+9}_{-7}$~AU) of the binary.}
    \label{fig:binary_orbit}
\end{figure}

To initialize our model we first used \texttt{orbitize!} \citep{2020AJ....159...89B} to fit the relative astrometry of \thisstar{}~B assuming system total system mass of $1.09+0.26=1.35~M_\odot$. \texttt{orbitize!} makes use of the Orbits For The Impatient (OFTI) Bayesian sampling method \citep{2017AJ....153..229B} that is well-suited to fitting the orbits of directly detected companions with orbital periods much longer than the observational span such as \thisstar{}~B. Next we extracted the posteriors from the \texttt{orbitize!} fit and calculated the corresponding $\chi^2$ value for a fit to the \emph{Hipparcos-Gaia} astrometry for each set of orbital parameters assuming stellar masses of $M_\mathrm{A}=1.09$~$M_\odot$, $M_\mathrm{B}=0.26$~$M_\odot$. This allowed us to identify a tightly constrained initial parameter space for most orbital parameters.

To model the orbit of \thisstar{}~B, we run a joint fit to the astrometry based on that of \citet{2019AJ....158..140B}. For modelling the \emph{Hipparcos-Gaia} astrometry we use the equations described in \citet{2021AJ....162...12V}, while the corresponding expressions for the relative astrometry can be found in \citet{1998A&AS..131..377P}. As in \citet{2021arXiv211103676V, 2021AJ....162...12V}, we resample the \emph{Hipparcos} and \emph{Gaia} proper motions using the observational epochs recorded in the Hipparcos Intermediate Astrometric Data for the former and the Gaia Observation Forecast Tool\footnote{\url{https://gaia.esac.esa.int/gost/}} for the latter. To explore the model parameter space we use the differential evolution MCMC sampler \texttt{edmcmc}\footnote{\url{https://github.com/avanderburg/edmcmc}} \citep{2021zndo...5599854V}.

A total of 11 parameters are used for the model: the system parallax $\varpi$, the primary mass $M_A$, the secondary mass $M_B$, the semi-major axis $a$, the eccentricity $e$ and argument of periastron $\omega$ parameterized as $\sqrt{e}\sin\omega$ and $\sqrt{e}\cos\omega$, the mean anomaly at an arbitrary reference epoch $\text{BJD}=2457000$, the orbital inclination $i$, the longitude of node $\Omega$, and finally two terms for the proper motion of the system barycenter.

The radial velocity trend generated by \thisstar{}~B is too small to be detected in the available data so we do not make use of RV data in the joint fit. The lack of radial velocity information in the fit results in the classical 180-degree degeneracy in the longitude of node $\Omega$ and argument of periastron $\omega$; following convention we report the solution with $\Omega$ in the range $[0,180]$~deg. The argument of periastron used in our model is that of the primary's orbit rather than that of the companion.

Of the 11 parameters used in the joint model, the parallax $\varpi$ was assigned a Gaussian prior of $17.061 \pm 0.037$~mas based on the \emph{Gaia}~EDR3 astrometric solution while the primary mass $M_A$ was given a Gaussian prior of $1.094 \pm 0.05$~$M_\odot$, with an inflated uncertainty as compared to the value in Table~\ref{tab:stellar} to avoid unduly biasing the model. Initial trials of the orbital fit were run without informed priors on the secondary mass $M_B$, however these runs tended to produce results skewed towards implausibly small values for this parameter ($M_B<0.1~M_\odot$), which in turn resulted in excessively broad distributions in the orbital parameters. It was therefore deemed prudent to adopt an informed prior of $M_B=0.26\pm0.04~M_\odot$ for the final model.

The results of our joint model for the orbit of \thisstar{}~B are presented in Table \ref{tab:binary}. Despite the span of observations being much shorter than the orbital period of the binary, we are able to robustly constrain most orbital parameters. We measure a semi-major axis of $a~=~56^{+9}_{-7}$~AU for \thisstar{}~B, approximately $\approx80\%$ larger than the projected separation, corresponding to an orbital period of $365^{+92}_{-69}$~yr. Our model shows a preference for relatively low orbital eccentricities ($0.25^{+0.22}_{-0.14}$, $e<0.61$ at $95\%$ confidence), resulting in a periastron separation of $43^{+11}_{-15}$~AU for the binary. We obtain a tightly constrained orbital inclination of $67.8^{+2.7}_{-2.9}$~degrees and a precise longitude of node of $20^{+11}_{-7}$~degrees, while our measurements of $\omega$ and the time of periastron for \thisstar{}~B are comparatively imprecise ($300^{+30}_{-80}$~deg, $2184^{+107}_{-74}$~CE).

Our fit to the astrometric data is shown in Figures \ref{fig:binary_fit} and \ref{fig:binary_orbit}. In the fit to the absolute astrometry it can be observed that the \emph{Hipparcos} proper motion measurement is too imprecise to significantly detect the astrometric signal. Conversely, the \emph{Gaia} and averaged \emph{Hipparcos-Gaia} proper motions are so precise that their uncertainties are not visible in the figure, and it is these measurements that drive the detection of the astrometric reflex signal of \thisstar{}~B. The importance of the $>10$-year timespan of relative astrometry is evident from Figures \ref{fig:binary_fit} and \ref{fig:binary_orbit}, as were it not for the detection of \thisstar{}~B in an archival NaCo observation in 2010 it would not be possible to derive such strong constraints on the binary orbit.

We note that the astrometric orbit solution is dependent on the priors we adopt for the system. For example, when the mass prior for \thisstar{}~B is removed, the orbital semi-major axis becomes degenerate with eccentricity. As such, future diffraction limited observations remain important in validating and refining the orbital parameters of the system. 

Based on our results, we predict the magnitude of the radial velocity trend on \thisstar{} generated by \thisstar{}~B to be $9.6\pm1.3~\ms~yr^{-1}$. This trend is not detectable in our RV data, but it is possible that future high-precision radial velocity measurements will be able to detect this acceleration. If so, the radial velocity information could then be used to improve the constraints on the orbit of \thisstar{}~B.

\section{Summary and Discussion}
\label{sec:discussion}

We report the discovery and statistical validation of a mini-Neptune around a bright young Sun-like star. \thisstarb{} is a $3.00_{-0.28}^{+0.32}\,R_\oplus$ planet in a 7.7\,day period orbit around its $V_\mathrm{mag}=8.31$ host star. Based on its kinematics, \thisstar{} can be placed in the AB Doradus moving group, with an age of $\sim 120$ Myr. The kinematics age is in agreement with that expected from the stellar rotation rate, lithium abundance, and X-ray emission intensity. 

The 600\,ppm transits of \thisstarb{} were identified at a signal-to-noise of 9.7 from \emph{TESS} Extended Mission Section 27 observations. However, the 2-day periodicity 2\% stellar variability led to the planetary signal slipping through the \emph{TESS} official planet selection processes. The detection of \thisstarb{} demonstrated the necessity of a dedicated search for planets around noisy, active young stars. Confirmation of transits came from the \emph{CHEOPS} mission. We obtained five orbits of observations with \emph{CHEOPS} to recover the transit of \thisstarb{}. Such follow-up would have been difficult to schedule with ground-based facilities due to the shallow transit. 

Figure~\ref{fig:terrpropagation} demonstrates the importance of CHEOPS follow-up in preserving the transit timing ephemeris. Without such observations, the timing derived from the three transits observed in the single sector \emph{TESS} observation would have eroded by $\sim 1$ hour per year. If a follow-up confirmation was not obtained within the first year after discovery, targeted transit observations would have been difficult to schedule and the recovery of such a small planet challenging \citep[see][]{2020AJ....159..219D}. 

\begin{figure*}
    \centering
    \begin{tabular}{cc}
        \includegraphics[width=0.45\linewidth]{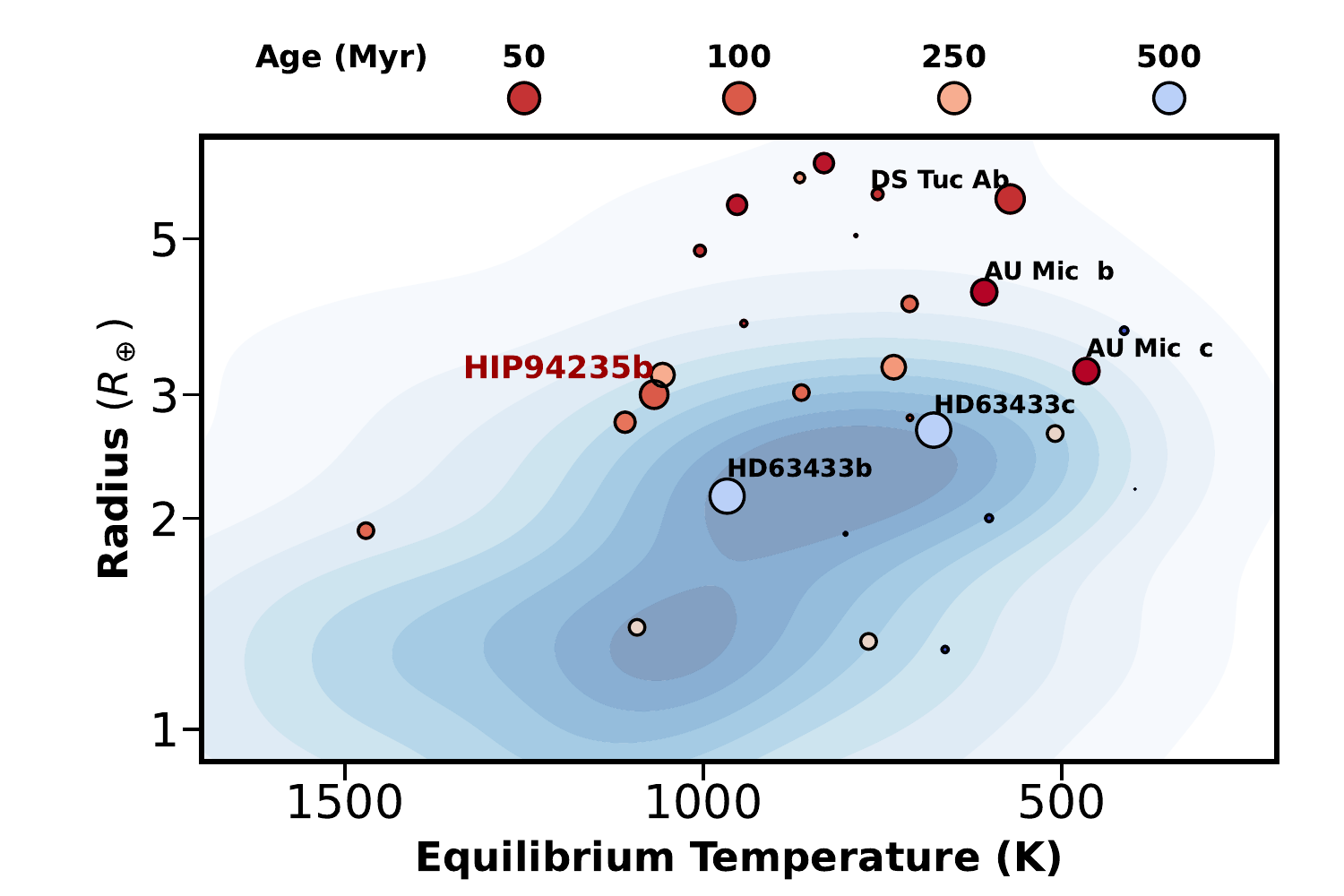} &
        \includegraphics[width=0.45\linewidth]{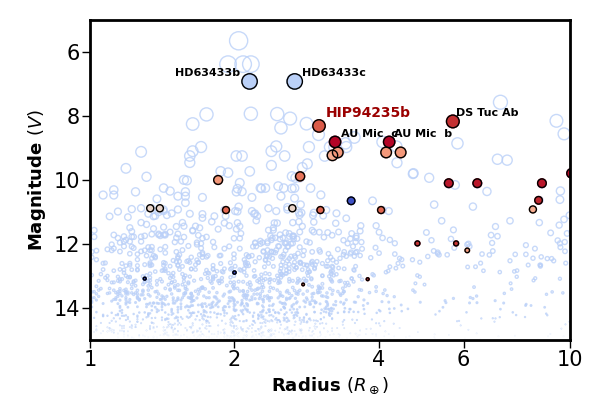} \\
    \end{tabular}
    \caption{\thisstarb{} amongst the distribution of small planets. The \textbf{left panel} shows the radius and equilibrium temperature distribution of planets from \citet{2017AJ....154..109F}. The sizes of each point mark the $V$ band magnitude of each host star, with the scale defined by the left panel. Planets around known young stars marked by the individual points. \thisstarb{} lies near the edge of the mini-Neptune desert. The \textbf{right panel} compares the planetary radius and host star magnitude of \thisstarb{} compared to other systems. Young planets are highlighted. The cyan points mark other transiting systems from the NASA Exoplanet Archive (January 2021). Young systems included in the figure are TOI-1227 \citep{2021arXiv211009531M}, V1298 Tau \citep{2019ApJ...885L..12D,2019AJ....158...79D}, K2-33 \citep{2016AJ....152...61M,2016Natur.534..658D}, DS Tuc A \citep{2019ApJ...880L..17N,2019A&A...630A..81B}, Kepler-63 \citep{2013ApJ...775...54S}, HIP 67522 \citep{2020AJ....160...33R}, HD63433 \citep{2020AJ....160..179M}, TOI-451 \citep{2021AJ....161...65N}, AU Mic \citep{2020Natur.582..497P}, TOI-251 \citep{2021AJ....161....2Z}, TOI-942 \citep{2021AJ....161....2Z,2021A&A...645A..71C}, K2-284 \citep{2018AJ....156..302D}, TOI-837 \citep{2020AJ....160..239B}, TOI-1098 \citep{2021arXiv210206066T}, K2-233 \citep{2018AJ....155..222D}, K2-77 \citep{2017MNRAS.464..850G}, K2-95, K2-100, K2-101, K2-102, K2-104, EPIC-211822797 \citep{2017AJ....153...64M}, TOI-1807, TOI-2076 \citep{2021AJ....162...54H}, Kepler 1627A \citep{2021arXiv211214776B}, TOI-1268 \citep{2022arXiv220112836D}.}
    \label{fig:periodradius}
\end{figure*}

\thisstarb{} is one of the smallest planets that has been found transiting a young star. Figure~\ref{fig:periodradius} shows the position of \thisstarb{} in the distribution of small planets around young stars. Most \emph{TESS} planets that transit young stars, including \thisstarb{}, have radii that place them above the radius valley delineated by \citet{2017AJ....154..109F} and \citet{2017ApJ...847...29O}. \thisstarb{} is at the edge of detectability for the \emph{TESS} light curve of \thisstar{}, and selection biases likely shape the current \emph{TESS} distribution of young planets. However, \citet{2017AJ....154..224R} found young clusters and associations surveyed by \emph{K2} hosted planets larger than the equivalent distribution about field stars. A similar holistic study of young stars surveyed by \emph{TESS} may elucidate the radius evolution timescale for young planets. 

Interestingly, despite continuous monitoring from the \emph{K2} mission of $\sim 1000$ Pleiades members, no confirmed planets have yet been found in the $125$\,Myr old cluster \citep[e.g.][]{2010MNRAS.408..475H,2017AJ....154..224R}. Systems such as TOI-451 \citep{2021AJ....161...65N} in the Pisces-Eridanus stream \citep{2019A&A...622L..13M,2019AJ....158...77C}, and \thisstar{} in AB Doradus, may help constrain the occurrence rates and radius properties of planets at the $\sim 100$\, Myr age range. These can help infer if the absence of planets in the Pleiades is due to detection biases, or if other astrophysical mechanisms may be at play. 

\begin{figure}
    \centering
    \includegraphics[width=1\linewidth]{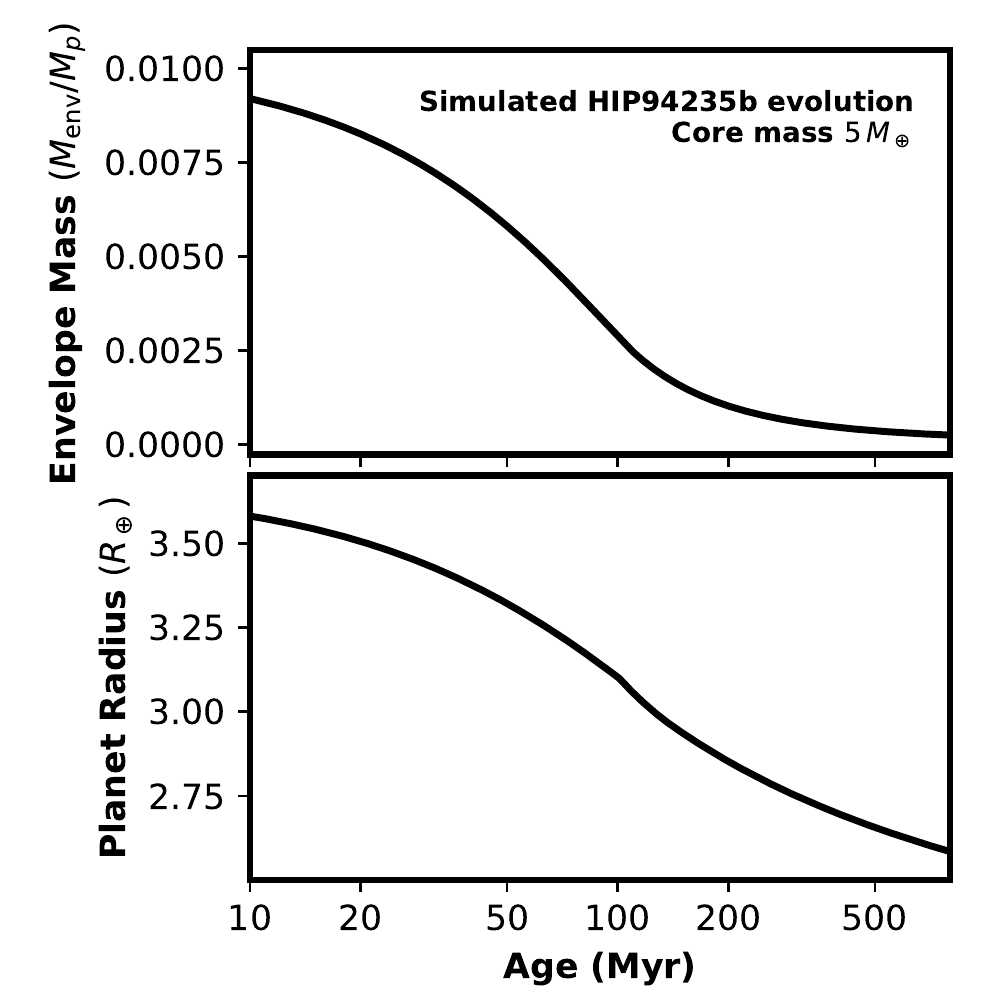}
    \caption{At 100 Myrs, \thisstar{} is undergoing run-away mass-loss evolution. The \textbf{top} shows the mass evolution of primordial envelope for the planet as per \citet{2017ApJ...847...29O}. Within the first $\sim 100$ Myr, \thisstarb{} is expected to lose most of its primordial envelope. The \textbf{bottom} panel shows the corresponding radius evolution expected for \thisstarb{} over the next few hundred million years. }
    \label{fig:massloss}
\end{figure}

Planets like \thisstarb{}, lying near the edge of the sub-Neptune valley, can provide key observational tests for the mechanisms of mass loss in young planets. \thisstarb{} and other recently discovered planets around young stars are subjected to significant high-energy radiation, which can be a dominant driver for rapid mass loss within the first hundreds of millions of years post formation \citep{2012MNRAS.425.2931O}. To estimate the mass evolution of \thisstarb{}, we adopt the analytical approach from \citet{2017ApJ...847...29O}. We find that we can replicate the current radius of \thisstarb{} via a planet model that has a high initial envelope mass fraction, and undergoes rapid mass loss with a time scale of $\sim$ 100-200\,Myr (Figure~\ref{fig:massloss}). \thisstarb{} has a current energy-limited mass-loss rate of $\sim 5\,M_\oplus\,\mathrm{Gyr}^{-1}$. At the end of this process, we expect the envelope mass fraction to reduce from 10\% to $\sim 1$\% of the total planet mass. Such mass and radius evolution is expected for many close-in Neptunes and super-Earths around Sun-like stars. 

\begin{figure*}
    \centering
    \includegraphics[width=0.7\linewidth]{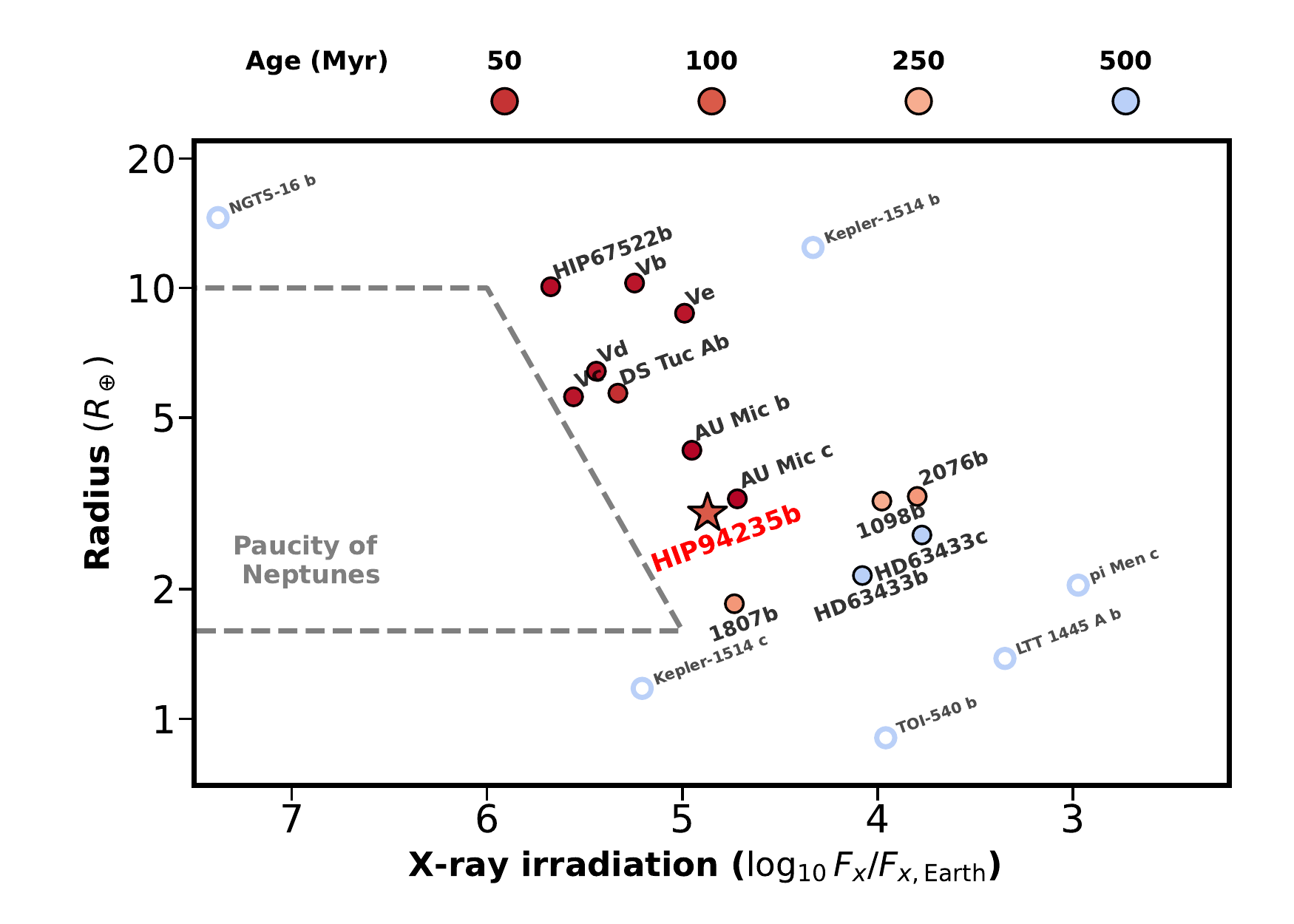}
    \caption{A large fraction of host stars that exhibit significant X-ray emission are known young stars. The X-ray irradiation received by all known transiting planets systems that have X-ray counterparts in the second ROSAT point source catalog \citep{2016AA...588A.103B} are shown. We see a paucity of planets with radii between $2-10\,R_\oplus$ in energetic environments. Older systems are marked by open light blue points, planets around known young stars or association and cluster members are marked by closed points, with older systems in light blue, younger systems in red. For brevity, names with `TOI' have been truncated to their TOI numbers only, and the V1298 Tau system are denoted by `Vb', `Vc', `Vd', `Ve'.}
    \label{fig:xray}
\end{figure*}

Figure~\ref{fig:xray} shows the X-ray irradiation experienced by \thisstarb{} compared to other systems about X-ray sources identified in the Second ROSAT Point Source Catalog \citep{2016AA...588A.103B}. To convert ROSAT count rates to fluxes, we adopt the calibration in \citet{1995ApJ...450..401F} and parallaxes made available from \citet{2018AA...616A...1G}. The majority of planetary systems around ROSAT X-ray sources are young, and the recent \emph{TESS} discoveries of nearby planet-hosting young stars are the most suitable targets for follow-up X-ray and UV observations. AU Mic \citep{2020Natur.582..497P} has the highest ROSAT count rate for any planet hosting star. Similarly, other nearby systems such as the planets around V1298 Tau \citep{2019AJ....158...79D,2019ApJ...885L..12D}, the ultra-short period super-Earth TOI-1807b \citep{2021AJ....162...54H}, DS Tuc Ab \citep{2019ApJ...880L..17N}, and \thisstarb{} mark the inner boundary of X-ray irradiation for small planets. The only planets around stars not identified as young in literature residing in more energetic environments are the hot Jupiter NGTS-16b \citep{2021MNRAS.504.6018T} and the Earth-sized inner planet in the Kepler-1514 system \citep{2021AJ....161..103D}. Neither have radii susceptible to significant modification by photoevaporation. Recent observations by \citet{2022AJ....163...68Z} detected the Lyman-alpha transit of HD63433c. Similar observations in the X-ray for young active stars have the potential of anchoring the photoevaporation models. 

Other avenues of mass loss are also possible. Atmospheric erosion from giant impact events can occur even more quickly, before compact super-Earth systems settle dynamically, acting on the tens of millions of years time scale post disk dispersal \citep[e.g.][]{2017MNRAS.470.1750I}. Core-powered mass loss may also reproduce the observed radius distribution independent of the host star irradiation \citep{2013ApJ...776....2L,2018MNRAS.476..759G}. By inferring the ages of \emph{Kepler} systems through numerous indicators, \citet{2021AJ....161..265D} found the radius gap may form at longer timescales. As such, it is unclear which mechanism dominates in shaping our current planet distribution. 

At 50\,AU, \thisstar{} is one of the tightest stellar binaries \citep{2021AJ....162..272S}. Like the DS Tuc AB system \citep{2019ApJ...880L..17N}, the orbit of the binary is aligned with that of the planetary orbit. This follows the trend from \citet{2022arXiv220200042C} that wide transiting planet-hosting binaries with separations between 100-700\,AU are preferentially found in edge-on orbits. \citet{2022arXiv220200042C} suggest that such trends may be due to the companions being formed from disk fragmentation, or the realignment of the inner disk by the perturbing outer companion. The efforts by \emph{TESS} follow-up teams to provide diffraction limited imaging of a majority of planet candidates, such as \thisstar{}, will help probe the continuation of this trend to $<100$\,AU separations.

The brightness of \thisstar{} makes the system suitable for follow-up atmospheric characterization with the next generation of space and ground-based facilities. Adopting the mass-radius relationship from \citet{2016ApJ...825...19W}, \thisstarb{} has a predicted mass of $11.2\pm1.4\,M_\oplus$, yielding a transmission spectroscopic metric of $96\pm25$. As such, \thisstarb{} ranks amongst the top dozen known planets between $1.6-4 \, R_\oplus$ in its suitability for follow-up transmission observations. In the era of JWST, we can compare the atmosphere of \thisstarb{} against planets of similar radii about older stars. We may find that young planets host primarily primordial atmospheres dominated by hydrogen and helium, with older planets hosting heavier water-rich atmospheres, or that some highly energetic environments may never allow secondary atmospheres to form. 
Orbital obliquities of young planets can help constrain the timescales of migration for planets that may have formed further out in their planetary systems. We expect a $\sim 10\,\mathrm{m\,s}^{-1}$ Rossiter-McLaughlin signal if \thisstarb{} is in a well aligned projected orbit. Stellar activity will be a limiting factor in achieving a secure detection of the spectroscopic transit, though past works have shown that this can be mitigated on transit-timescales due to the smoothly varying nature of the rotational modulated velocity noise \citep[e.g.][]{2020A&A...643A..25P,2021A&A...650A..66B}. 

\acknowledgements  
We respectfully acknowledge the traditional custodians of all lands throughout Australia, and recognise their continued cultural and spiritual connection to the land, waterways, cosmos, and community. We pay our deepest respects to all Elders, ancestors and descendants of the Giabal, Jarowair, and Kambuwal nations, upon whose lands the MINERVA-Australis facility at Mt Kent is situated.
GZ thanks the support of the ARC DECRA program DE210101893.
CW and GZ thank the support of the TESS Guest Investigator Program G03007.
CH thanks the support of the ARC DECRA program DE200101840.
EG gratefully acknowledges support from the David and Claudia Harding Foundation in the form of a Winton Exoplanet Fellowship.
CHEOPS is an ESA mission in partnership with Switzerland with important contributions to the payload and the ground segment from Austria, Belgium, France, Germany, Hungary, Italy, Portugal, Spain, Sweden and the United Kingdom. We thank support from the CHEOPS GO Programme and Science Operations Centre for help in the preparation and analysis of the CHEOPS observations.
This research has used data from the CTIO/SMARTS 1.5m telescope, which is
operated as part of the SMARTS Consortium by \href{http://secure-web.cisco.com/1TL5nionOJJUGi7T0X_YvX7RLRwbVQl20QG7s4LKeK1vpFY8M3UHYMuONVvV2D2hxli_pMi4YkHdTYel4ogZ3sJWN4axM8-5IsyCIPeIj7BfVIBOvp9a8iRKv2IM-wTBpjGA3xxZcH5lT4FNKBIoEstyJEEyUYzEKbDQyL4T1LQSiukl5eTarVlkS9YJbHf_HrjiuXV1gM1uXr7gdIdCbZg4CfJa_N8Qw38oz0KhpJ74RZ0rIcyg3XKCc6-HCDjlBrMtX3cpMKa1Kcya1SxY0UxXY0WkwM0zGeXYUYfbkp1Ce6jIBY8Evcz-YcyODRE4QWMlPqSDV66bKv5F1R3-RrkcH91Y7INyFOP6qJfGJKLRFJT-KNphpqmNc4Pf7zLVOIBjCEKsANmt1XTtzQN5AIPwKf-F1qd4b6KCZrqjHZIA/http\%3A\%2F\%2Fwww.recons.org}{RECONS}
This study was based in part on observations made using the Las Cumbres Observatory global telescope network, using time allocated by the National Science Foundation’s NOIRLab (NOIRLab Prop. ID NOAO2021A-009; principal investigator: J. Hartman). Some of the observations reported in this paper were obtained with the Southern African Large Telescope (SALT).
Some of the observations in the paper made use of the High-Resolution Imaging instrument Zorro obtained under Gemini LLP Proposal Number: GN/S-2021A-LP-105. Zorro was funded by the NASA Exoplanet Exploration Program and built at the NASA Ames Research Center by Steve B. Howell, Nic Scott, Elliott P. Horch, and Emmett Quigley. Zorro was mounted on the Gemini North (and/or South) telescope of the international Gemini Observatory, a program of NSF’s OIR Lab, which is managed by the Association of Universities for Research in Astronomy (AURA) under a cooperative agreement with the National Science Foundation. on behalf of the Gemini partnership: the National Science Foundation (United States), National Research Council (Canada), Agencia Nacional de Investigación y Desarrollo (Chile), Ministerio de Ciencia, Tecnología e Innovación (Argentina), Ministério da Ciência, Tecnologia, Inovações e Comunicações (Brazil), and Korea Astronomy and Space Science Institute (Republic of Korea).
This research has made use of the NASA Exoplanet
Archive, which is operated by the California Institute of Technology,
under contract with the National Aeronautics and Space Administration
under the Exoplanet Exploration Program. 
Funding for the TESS mission is provided by NASA's Science Mission directorate. We acknowledge the use of public TESS Alert data from pipelines at the TESS Science Office and at the TESS Science Processing Operations Center. This research has made use of the Exoplanet Follow-up Observation Program website, which is operated by the California Institute of Technology, under contract with the National Aeronautics and Space Administration under the Exoplanet Exploration Program. This paper includes data collected by the TESS mission, which are publicly available from the Mikulski Archive for Space Telescopes (MAST).
Resources supporting this work were provided by the NASA High-End Computing (HEC) Program through the NASA Advanced Supercomputing (NAS) Division at Ames Research Center for the production of the SPOC data products.
MINERVA-Australis is supported by Australian Research Council LIEF Grant LE160100001, Discovery Grants DP180100972 and DP220100365, Mount Cuba Astronomical Foundation, and institutional partners University of Southern Queensland, UNSW Sydney, MIT, Nanjing University, George Mason University, University of Louisville, University of California Riverside, University of Florida, and The University of Texas at Austin.
D. D. acknowledges support from the TESS Guest Investigator Program grants 80NSSC21K0108 and 80NSSC22K0185.

\facility{TESS, CHEOPS, Exoplanet Archive, CTIO 1.5\,m, LCOGT, Gemini:Zorro, SALT, MINERVA-Australis}
\software{emcee \citep{2013PASP..125..306F}, batman \citep{2015PASP..127.1161K}, astropy \citep{2018AJ....156..123A}, PyAstronomy \citep{pya}, comove \citep{2021arXiv210206066T}}
\bibliographystyle{apj}
\bibliography{refs}

\end{document}